\documentclass[%
aps,
prd,
twocolumn,
superscriptaddress,
preprintnumbers,
nofootinbib,
amsmath,amssymb,
]{revtex4-2}

\usepackage{style}
\usepackage{aas_macros}

\begin{document}

\preprint{Imperial--TP--2026--AM--02}
\preprint{DESY-26-075}

\title{Probing Quadratically Coupled Ultralight Dark Matter 
\\
with the Laser Interferometer Space Antenna}

\author{Xucheng Gan\,\orcidlink{0000-0003-2834-7498}}
\affiliation{Deutsches Elektronen-Synchrotron DESY, Notkestr. 85, 22607 Hamburg, Germany}

\author{Hyungjin Kim\,\orcidlink{0000-0002-8843-7690}}
\affiliation{Universit\'e Paris-Saclay, CNRS, CEA, Institut de Physique Th\'eorique, 91191 Gif-sur-Yvette, France}

\author{Anna-Malin Lemke\,\orcidlink{0009-0005-3568-3336}}
\affiliation{Deutsches Elektronen-Synchrotron DESY, Notkestr. 85, 22607 Hamburg, Germany}

\author{Andrea Mitridate\,\orcidlink{0000-0003-2898-5844}}
\affiliation{Abdus Salam Centre for Theoretical Physics, Imperial College, London, SW7 2AZ, UK}

\begin{abstract}
Ultralight dark matter can interact with Standard Model particles via gravitational and non-gravitational interactions. Through such interactions, it can leave distinctive signals in gravitational-wave experiments. In this work, we investigate signals induced by ultralight dark matter quadratically coupled to the Standard Model in the future space-borne gravitational-wave detector, the Laser Interferometer Space Antenna (LISA). Due to the quadratic nature of the coupling, dark matter signals appear at two distinct frequencies: the frequency corresponding to twice the dark matter mass, and frequencies below the typical dark matter kinetic energy. We analyze both contributions and show that LISA can surpass current constraints from terrestrial and astrophysical probes in certain mass ranges. We also find that dark matter signals in LISA are free from screening effects which significantly limit the sensitivity of terrestrial experiments.
\end{abstract}

\maketitle
\tableofcontents

\bigskip

\section{Introduction}
The Laser Interferometer Space Antenna (LISA) mission was approved by the European Space Agency in January 2024 and is expected to be launched in 2035~\cite{LISA:2017pwj, LISA:2024hlh}. LISA will probe gravitational waves (GWs), emitted from a variety of sources, including massive black hole binaries~\cite{Hughes:2001ya, Sesana:2010wy, Klein:2015hvg}, extreme mass ratio inspirals~\cite{Amaro-Seoane:2007osp, Babak:2017tow, Berry:2019wgg}, compact galactic binaries~\cite{Nelemans:2001hp, Nissanke:2012eh, Korol:2017qcx, Kupfer:2018jee, Finch:2022prg}, and potentially cosmological sources~\cite{Caprini:2015zlo, Auclair:2019wcv, Caprini:2019egz, LISACosmologyWorkingGroup:2022jok, Caprini:2024hue}.

Beyond gravitational waves, LISA could probe dark matter (DM) near the solar system. As DM scatters off the test masses, it could leave detectable signals in the detector output. Promising targets include primordial black holes~\cite{Adams:2004pk,Seto:2004zu}, dark clumps~\cite{Baum:2022duc}, and super-Planckian composite DM candidates~\cite{Hall:2016usm,Du:2023dhk} -- all of which are heavy DM candidates, expected to generate a larger momentum transfer to the test mass compared to light DM candidates.

Interestingly, LISA is also sensitive to ultralight dark matter (ULDM) candidates. Ultralight dark matter refers to bosonic DM with mass below an eV. Such candidates can be described as coherent classical waves, oscillating at a frequency determined by their mass, with correlation lengths extending over astrophysical scales. These coherent oscillations perturb the propagation of laser beams and the motion of test masses, leaving distinctive signals in the detector data streams. Previous studies have forecast the sensitivity of LISA and other GW detectors to such coherent signals, considering both gravitational~\cite{Aoki:2016kwl, Kim:2023pkx, Yu:2024enm, Badurina:2025xwl, Foster:2025nzf} and non-gravitational couplings~\cite{Pierce:2018xmy, Morisaki:2018htj, Badurina:2019hst, Grote:2019uvn, Morisaki:2020gui, MAGIS-100:2021etm, Frerick:2023xnf, Yu:2023iog, Yao:2024fie, Baynham:2025pzm, Jiang:2026haa}.

When DM couples quadratically to the Standard Model (SM), its signal exhibits additional stochastic oscillations at frequencies below the typical DM kinetic energy. This is a generic feature of quadratic interactions: a primary example is the gravitational coupling between DM and SM particles, since the energy-momentum tensor is quadratic in the DM field. This has motivated studies investigating the sensitivity of several GW experiments to stochastic signals induced by the ULDM gravitational coupling, including interferometers~\cite{Kim:2023pkx}, pulsar timing arrays (PTAs)~\cite{Kim:2023kyy, Eberhardt:2024ocm, Dror:2025nvg, Gan:2025icr}, astrometric observations~\cite{Kim:2024xcr}, lunar laser range~\cite{Foster:2025csl}, and interplanetary radio range~\cite{Frerick:2026kzq}. The studies showed that such experiments can place the strongest direct limits on the abundance of ULDM candidates near the solar system -- a quantity that has no direct observational evidence at the moment.

To date, studies of stochastic ULDM signals have focused primarily on gravitational interactions. However, ULDM could also have quadratic non-gravitational couplings to the SM. In this work, we investigate stochastic and coherent signals from non-gravitational quadratic interactions and forecast the sensitivity of LISA to such signals.

We find that LISA has a distinctive advantage over other existing searches for non-gravitational quadratic interactions. Quadratic interactions tend to distort the DM field profile near objects of finite density, often suppressing the ULDM field near the detector and reducing the DM signals. This {\it screening effect} significantly affects the sensitivity of existing probes of ULDM quadratic interactions, such as pulsar timing arrays~\cite{Gan:2025icr}, atomic clocks and spectroscopy~\cite{Arvanitaki:2014faa, Hees:2016gop, Kennedy:2020bac, BACON:2020ubh, Oswald:2021vtc, Sherrill:2023zah, Filzinger:2023zrs, Arakawa:2026mls}, and equivalence principle tests~\cite{Hees:2018fpg, MICROSCOPE:2019jix, MICROSCOPE:2022doy,Bouley:2026frx,Brzeminski:2026pgz}, as they all operate on or near Earth's surface. In contrast, we find that LISA is unaffected by this screening effect because it operates far from dense environments and employs much smaller test masses. Combined with LISA's expected sensitivity, this fact will allow LISA to probe ULDM quadratic couplings beyond the limits set by other probes for masses above $\sim 10^{-14}\,{\rm eV}$. 

We also revisit the analysis of stochastic signals produced by ULDM gravitational interactions. The sensitivity of LISA to such signals was already estimated in Ref.~\cite{Kim:2023pkx}, but the estimate was obtained by assuming that two LISA-like constellations would operate at the same time, such that their data could be cross-correlated. Whether such an analysis will be feasible depends on the operational timeline of future space-borne gravitational-wave missions, such as TianQin~\cite{TianQin:2015yph,TianQin:2020hid}, Taiji~\cite{Hu:2017mde,Gong:2021gvw}, and DECIGO~\cite{Kawamura:2011zz}. Here we show that even a single LISA constellation can place upper limits on the DM abundance near the solar system comparable to those of Ref.~\cite{Kim:2023pkx}.

This work is organized as follows. In Sec.~\ref{sec:uldm}, we summarize the basic properties of ULDM. In Sec.~\ref{sec:uldm_signal}, we discuss ULDM signals in LISA by deriving ULDM-induced single-link signals and then generalizing them to arbitrary time-delay interferometer (TDI) variables. In Sec.~\ref{sec:analysis}, we summarize the analyses that we perform to derive the projected sensitivities. The results are presented in Sec.~\ref{sec:results}, and discussed in Sec.~\ref{sec:discussion}. Finally, we conclude in Sec.~\ref{sec:conclusion}.

\section{Ultralight Dark Matter}\label{sec:uldm}
In this section, we introduce the conventions used to describe the DM and its interactions. The Lagrangian is given by
\begin{equation}
    \mathcal{L} = \mathcal{L}_0 + \mathcal{L}_{\rm int}\,,
\end{equation}
where the free and interaction Lagrangians are given by
\begin{align}
    \mathcal{L}_0 
    &= 
    \frac{1}{2} g^{\mu\nu} \partial_\mu \phi \partial_\nu \phi
    - \frac{1}{2} m^2_\phi \phi^2
    \,,
\\
\mathcal{L}_{\rm int}
&=
-
\frac{\phi^2}{2\Mpl^2} \sum_i d_i \mathcal{O}_{\rm SM}^{(i)}\ . 
\label{L_int}
\end{align}
Here, $m_\phi$ is the DM mass, $M_{\rm pl} = (4\pi G)^{-1/2}= 3.4 \times 10^{18}\,{\rm GeV}$ is the Planck mass, and $\mathcal{O}_{\rm SM}^{(i)}$ are the relevant SM operators. We choose the conformal Newtonian gauge to parameterize the metric:
\begin{align}\label{metric}
ds^2 = \big(1 + 2 \Phi \big) dt^2 - \big( 1 - 2 \Psi\big)dx^2\, .
\end{align}
Couplings between the ULDM field and SM operators induce a dependence of fundamental constants on the background field configuration. For example, a coupling to the SM operator $O_{\rm SM}^{(m_e)}=m_e\bar{\psi}_e \psi_e$ induces fluctuations in the electron mass, given by
\begin{equation}
    \frac{\delta m_e}{m_e}=d_{m_e}\frac{\phi^2}{2 M_{\rm pl}^2}\,.
\end{equation}
These fluctuations in fundamental constants propagate to the LISA test masses, making their values depend on the local ULDM field configuration and giving rise to some of the signals discussed in the next section.

We model the DM field as a classical Gaussian random field that we expand in plane waves as\footnote{The plane wave is not the solution of the equation of motion due to the gravitational potential of the Sun. An actual eigenfunction of the system is given by the confluent hypergeometric function, but the quantitative difference between the plane wave approximation and the full solution remains at the few percent level for a generic halo dark matter~\cite{Kim:2021yyo}. 
}
\begin{equation}\label{plane_wave_expansion}
    \phi(t, \boldsymbol x)
    \approx \frac{1}{\sqrt{2m_\phi V}} \sum_{\boldsymbol k}
    \left[
    a_{\boldsymbol k} e^{-i k \cdot x}
    + a_{\boldsymbol k}^{*} e^{i k \cdot x}
    \right]\,,
\end{equation}
where the sum runs over the momentum modes, and $\{a_{\boldsymbol k}\}$ is a set of complex random variables. The complex coefficients, $a_{\boldsymbol{k}}$, can be decomposed as $a_{\boldsymbol k} = r_{\boldsymbol k} e^{i\theta_{\boldsymbol k}}$, with the real amplitude, $r_{\boldsymbol k}$, and phase, $\theta_{\boldsymbol k}$, following a Rayleigh and uniform distribution, respectively~\cite{Derevianko:2016vpm, Foster:2017hbq, Centers:2019dyn, Kim:2021yyo, Cheong:2024ose}:
\begin{align}
    \mathcal{P}(r_{\boldsymbol k}) &= 
    \frac{2 r_{\boldsymbol k}}{f_{\boldsymbol k}} 
    \exp\left( - \frac{r_{\boldsymbol k}^2 }{ f_{\boldsymbol k} } \right)\,, \label{eq:p_r}
    \\
    \mathcal{P}(\theta_{\boldsymbol k}) &= \frac{1}{2\pi}\,. \label{eq:p_theta}
\end{align}
In the continuum limit, $f_{\boldsymbol k}$ can be interpreted as the DM momentum distribution, which we choose to be an isotropic Maxwell-Boltzmann distribution:
\begin{equation}
    f( \boldsymbol{k} )
    = \frac{(2\pi)^{\frac32}}{(m_\phi \sigma)^3} 
    \frac{\bar\rho}{m_\phi}
    \exp\left( - \frac{|\boldsymbol{k}|^2}{2 m_\phi^2 \sigma^2} \right)\,,
\end{equation}
with $\sigma \approx 160\,{\rm km/s}$ from the Standard Halo Model~\cite{Freese:2012xd}. Here, we have normalized the scale parameter, $f_{\boldsymbol k}$, such that $\bar \rho = (m_\phi / V) \sum_{\boldsymbol k} f_{\boldsymbol k}= m_\phi\int [d^3\boldsymbol{k}/(2\pi)^3] f({\boldsymbol k})$ with $\bar \rho$ being the mean dark matter density. The mean velocity is ignored for simplicity. 

The statistical properties given in Eqs.~\eqref{eq:p_r}--\eqref{eq:p_theta} can then be used to compute the ensemble average of any function of the DM field. Specifically, we can use them to derive the power spectrum of the quadratic field operator, $P_{\phi^2}(k)$, defined as
\begin{equation}
    \langle
    \widetilde{\phi^2}(k) 
    \widetilde{\phi^2}^*(k') 
    \rangle
    = (2\pi)^4 \delta^{(4)}(k - k') P_{\phi^2}(k)\,,
    \label{phi2_twoPoint}
\end{equation}
where $k=(\omega,\boldsymbol{k})$ is the four-momentum, and $\widetilde{\phi^2}(k) = \int d^4 x \, e^{i k \cdot x} \phi^2(x)$ is the Fourier transform of the quadratic field operator $\phi^2(x)$. Since we are not interested in the zero-mode signal, any quadratic operator in the following will be understood as a fluctuation around the mean value, i.e. $\phi^2 \to \phi^2 - \langle \phi^2 \rangle$. 

The power spectrum consists of two contributions at widely separated frequencies, and it can be decomposed as~\cite{Gan:2025icr, Kim:2023pvt}
\begin{equation}\label{eq:phi_ps}
    P_{\phi^2}(k) = 
    P^{\rm fast}_{\phi^2}(k)
    + 
    P^{\rm slow}_{\phi^2}(k)\,,
\end{equation}
where both spectra are one-sided and given by
\begin{align}
    \label{eq:phi_fast}
    P_{\phi^2}^{\rm fast}(k) 
    &= 
    \frac{2\pi^2\bar\rho^2}{m_\phi^8\sigma^5}
    e^{-\frac{\omega-2m_\phi}{m_\phi\sigma^2}}
    \sqrt{\frac{\omega-2m_\phi}{m_\phi\sigma^2}-\frac{|\bm{k}|^2}{4 m_\phi^2 \sigma^2}}\,,
\\
    \label{eq:phi_slow}
    P_{\phi^2}^{\rm slow}(k)
    &= 
    \frac{2\pi^2 \bar\rho^2}{|\bm{k}| m_\phi^7 \sigma^4}
    \exp\bigg[
    -\frac{|\bm{k}|^2}{4m_\phi^2\sigma^2} - \frac{\omega^2}{\sigma^2|\bm{k}|^2}
    \bigg]\,.
\end{align}
We label each quantity with fast and slow mode spectrum since each represents the fluctuations of $\phi^2$ at $\omega = 2m_\phi$ and $\omega \lesssim m_\phi \sigma^2$. The fast mode power spectrum is valid only for $\omega >0$; the spectrum for $\omega < 0$ can be obtained by taking $\omega \to -\omega$ from the expression above. The fast and slow mode spectra describe the coherent and stochastic oscillations of the DM field, respectively. For this reason, in the following we use the terms {\it fast mode signal} and {\it coherent signal} interchangeably, and likewise for the {\it slow mode signal} and {\it stochastic signal}. In the following section, we will show that the dark matter signal in LISA can be completely parameterized in terms of the power spectrum of the quadratic operator. 

\section{ULDM signal in LISA}\label{sec:uldm_signal}
In this section, we derive the signals in the LISA data stream induced by quadratic couplings of ultralight dark matter. We begin by examining the response of a single-link detector (Sec.~\ref{subsec:single_link}) and then generalize the result to a realistic time-delay interferometry setup (Sec.~\ref{subsec:tdi}). 

\begin{figure}
\centering
\includegraphics[width=0.45\textwidth]{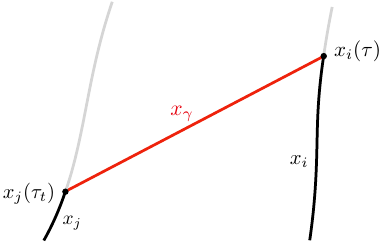}
\caption{Spacetime diagram showing the worldlines of two free-falling stations (black curves) and the photon path $x_\gamma^\mu(\lambda)$ connecting them (red curve). Light is emitted from station $j$ at $x^\mu_j(\tau_t)$ and received at station $i$ at $x^\mu_i(\tau)$.}
\label{fig:worldlines}
\end{figure}

\subsection{Single-Link Detector}\label{subsec:single_link}
Consider a detector consisting of two stations in free fall, moving along worldlines $x^\mu_{i,j}(\tau_{i,j})$ parametrized by their respective proper times, $\tau_{i,j}$, as illustrated in Fig.~\ref{fig:worldlines}. Suppose that laser light is transmitted from the $j^{\rm th}$ to the $i^{\rm th}$ station. Upon arrival, this light is mixed with a local laser, and the phase difference 
\begin{align}
s_{ij}(t) = s_i(t) - s_j(t)
\end{align}
is measured by a phasemeter, where $s_{i,j}(t)$ are the phases of the two lasers at the time of reception, $t$. This phase difference is the key observable in our discussion of ULDM searches in LISA. The purpose of the rest of this section is to characterize the DM signals encoded in this observable. 

The phase difference measured at the receiving station at time $t=x^0_i(\tau)$ is given by 
\begin{align}
    s_{ij}(t) =& \, \omega_{\rm het} t + \omega_{L,j}\Delta t 
    \nonumber\\
    &
    + \omega_{L,i} \int^\tau d\tau_i\,\Delta_{L,i} 
    - \omega_{L,j} \int^{\tau_t} d\tau_j\,\Delta_{L,j}\,,  
    \label{eq:s_ij_full}
\end{align}
where $\omega_{\rm het} = \omega_{L,i} - \omega_{L,j}$ is a heterodyne beat frequency, $\omega_{L,i}$ is the laser frequency in the $i^{\rm th}$ station, $\Delta t$ is the light travel time between the two stations, and the integrals in the last two terms should be carried out along the worldline of each test mass. In addition, we introduce
\begin{equation}
\Delta_{L,i}=
\frac{\delta\omega_{L,i}}{\omega_{L,i}} 
+ \Phi(x_i(\tau_i)) - \frac{v_i^2}{2},
\label{laser_phase}
\end{equation}
where $\delta\omega_{L,i}/ \omega_{L,i}$ describes the fractional frequency uncertainty of the laser and $v_i$ is the velocity of the $i^{\rm th}$ station. The factor $\Phi - v_i^2/2$, arising from the difference between proper time and coordinate time, acts as additional laser frequency noise. The intrinsic laser frequency noise, $\delta\omega_{L,i}/\omega_{L,i}$, would largely obscure any GW signal in LISA if not properly accounted for. For this reason, as we discuss in the following section, different single-link measurements are combined to form time-delay interferometry (TDI) variables in which the dominant laser frequency noise cancels out. Since the proper time corrections in $\Delta_{L,i}$ behave as additional laser frequency noise, they are also suppressed by TDI. We therefore neglect this contribution in what follows. 

The DM signal is therefore dominated by fluctuations in the light travel time. It can be obtained by integrating the photon geodesic equation between the two stations:
\begin{equation}\label{eq:time_delay}
    \Delta t(t)
    \!\approx \!
    L_{ij} + \hat{\boldsymbol n}_{ij}\!\cdot\! 
    \big[ \boldsymbol{\delta x}_i(t) - \boldsymbol{\delta x}_j(t-L_{ij}) \big]
    \!- \omega_L\!\! \int^{\lambda}_{\lambda_t} \!\! d\lambda' (\Phi + \Psi)\,,    
\end{equation}
where $L_{ij}$ is the unperturbed distance between the stations, $\hat{\boldsymbol{n}}_{ij}$ is a unit vector pointing from the $j^{\rm th}$ to the $i^{\rm th}$ station, $\boldsymbol{\delta x}_i$  is the perturbation of the $i^{\rm th}$ station from its unperturbed position $\boldsymbol{x}_{i}$, and the integral is carried out along the photon geodesic. In the last term, we assume that the laser frequency is the same for all stations and is denoted by $\omega_L$ for convenience. The derivation of the expressions \eqref{eq:s_ij_full}--\eqref{eq:time_delay} is detailed in App.~\ref{app:sig_derivation}.

In Eq.~\eqref{eq:time_delay}, ULDM induces two time-dependent contributions to the light travel time: one through the perturbation of the station positions [second term on the right-hand side of Eq.~\eqref{eq:time_delay}], and the other through the Shapiro time delay arising from the time-dependent potential along the photon path [last term on the right-hand side of Eq.~\eqref{eq:time_delay}]. In what follows, we discuss the relative contribution of these two signals and connect them to the ULDM properties.

\smallskip

\emph{\textbf{Acceleration signal}} -- The phase shift due to the perturbation of the test mass position by ULDM is given by
\begin{equation}\label{eq:acc_signal}
    s_{ij}(t) = \omega_L \hat{\boldsymbol n}_{ij} \cdot 
    \big[
    \boldsymbol {\delta x}_i(t) - \boldsymbol{\delta x}_j(t - L_{ij}) 
    \big]\,,
\end{equation}
where the perturbation $\boldsymbol{\delta x}(t)$ evolves according to
\begin{equation}\label{eq:acc}
    \boldsymbol{a}_i(t) = \ddot{\boldsymbol{\delta x}_i}(t)=\begin{cases}
    -\boldsymbol\nabla\Phi
    & \textrm{grav.}
    \\
    -\dfrac{g}{2\Mpl^2}\boldsymbol\nabla\phi^2\,,
    &
    \textrm{non-grav.}
    \end{cases}.
\end{equation}
The gravitational acceleration is sourced by the Newtonian potential $\Phi$. The non-gravitational acceleration arises because direct couplings between the DM field and the SM render the mass of the test masses dependent on the local field configuration, $M\to M + \delta M[\phi^2]$: as the field fluctuates, so does the test mass, producing an additional acceleration proportional to $\boldsymbol{\nabla}\phi^2$. The dimensionless parameter $g$ characterizes the effective coupling strength between the ULDM field and the test mass; its relation to the Lagrangian couplings $d_i$, Eq.~\eqref{L_int}, is given in Sec.~\ref{sec:results}.

Below, we characterize the ULDM signals via their respective power spectra. As an intermediate step, we compute the Fourier component of the single-link signal. By using Eqs.~\eqref{eq:acc_signal}--\eqref{eq:acc}, we find the Fourier transform of the signal induced by gravitational interactions as
\begin{equation}\label{eq:one_way_pert_acc}
    \tilde s_{ij}(f) 
    = 
    i \omega_L
    \int\frac{d^3\boldsymbol{k}}{(2\pi)^3}
    \frac{\boldsymbol k \cdot \hat{\boldsymbol n}_{ij}}{\omega^2}
    {\cal R}_{ij}(k)
    \tilde \Phi(k) e^{i \boldsymbol k \cdot \boldsymbol x_i} \,,
\end{equation}
where $\omega = 2 \pi f $ is an angular frequency and ${\cal R}_{ij}(k)$ is the single-link response function,
\begin{equation}
    {\cal R}_{ij}(k) \equiv  1 - e^{i \omega L_{ij} - i {\bm k}\cdot {\bm L}_{ij}}\,. 
\end{equation}
Here $\tilde \Phi(k) = \tilde \Phi(f,\boldsymbol k)$ is the four-dimensional Fourier transform of the potential, ${\bm L}_{ij} = {\bm x}_i - {\bm x}_j$, and we have assumed that the unit vector $\hat{\boldsymbol n}_{ij}$ is time-independent. The Fourier components of the acceleration signal produced by non-gravitational interactions take the same form as Eq.~\eqref{eq:one_way_pert_acc}, with $\tilde\Phi$ replaced by $g\,\widetilde{\phi^2}/(2M_{\rm pl}^2)$.

\smallskip

\emph{\textbf{Shapiro signal}} -- By perturbing the metric along the photon geodesic, the ULDM field can also induce a Shapiro time delay between the stations. The Shapiro delay signal is given by
\begin{align}
    s_{ij}(t) 
    &= - \omega_L^2 \int^\lambda_{\lambda_t} d\lambda' \, 
    \big[
         \Phi(x_\gamma(\lambda')) + \Psi(x_\gamma(\lambda'))
    \big]
    \,,
\\
    \tilde s_{ij}(f) 
    &= 
    -i \omega_L 
    \int\frac{d^3\boldsymbol{k}}{(2\pi)^3}
    \frac{\tilde \Phi(k) + \tilde \Psi(k)}{\omega -  \boldsymbol k \cdot \hat {\boldsymbol n}_{ij} } 
    {\cal R}_{ij}(k)
    e^{i \boldsymbol k \cdot \boldsymbol x_i} \,,
    \label{eq:one_way_shapiro}
\end{align}
where the Fourier transformation is obtained by performing the integration over the unperturbed photon worldline, $x_\gamma^\mu(\lambda) = x_\gamma^\mu(\lambda_t) + (\lambda - \lambda_t) k_\gamma^\mu$, with an unperturbed photon four momentum $k^\mu_\gamma = \omega_L(1, \hat{\boldsymbol n}_{ij})$ and a boundary condition $x_\gamma^\mu(\lambda_t) = x_j^\mu(\tau_t)$. 

\smallskip

We can already compare the relative importance of the two signals by analyzing the integrand of the Fourier components of each signal. For stochastic fluctuations, the momentum integrals in Eqs.~\eqref{eq:one_way_pert_acc} and \eqref{eq:one_way_shapiro} have support at $\omega \lesssim m_\phi \sigma^2$ and $k \sim m_\phi \sigma$. Since $\tilde \Phi(k) \approx \tilde\Psi(k)$ for slow mode fluctuations, the Shapiro contribution is suppressed by a factor of $\omega/k \sim \sigma$ relative to the gravitational acceleration signal.

For coherent fluctuations, the integrand has narrow support around $\omega = 2m_\phi$, while $k \sim m_\phi \sigma$. A direct comparison of Eqs.~\eqref{eq:one_way_pert_acc} and \eqref{eq:one_way_shapiro} would suggest that the Shapiro signal is $\mathcal{O}(1/\sigma)$ larger than the gravitational acceleration signal. 
Although this estimate holds for a single-link measurement, the two signals turn out to be of the same order in the TDI variables that we use to forecast LISA's sensitivity to ULDM signals.

To illustrate this, we first expand the Shapiro delay expression in the small $k/\omega$ limit. We find 
\begin{align}
\!\!\! 
\tilde s_{ij}(f) \approx - \frac{i\omega_L}{\omega}
\int \frac{d^3 \boldsymbol k}{(2\pi)^3} 
{\cal R}_{ij}(k) \big( \tilde\Phi(k) + \tilde\Psi(k) \big) e^{i \boldsymbol k \cdot \boldsymbol x_i},
\label{shapiro_leading}
\end{align}
If we further assume that the arm-lengths are equal, i.e. $L_{ij} = L$,  we see from the expression above that the leading order Shapiro signal in the $k/\omega$ expansion is identical across all links in the LISA constellation. 
Since TDI observables are constructed by taking differences between signals of different links, this leading-order contribution cancels out~\cite{Kim:2023pkx}.\footnote{The equal-arm approximation is not required for this cancellation. We project the sensitivity of LISA using 1.5 generation TDI variables, which are designed to cancel laser phase noise even when the arm-lengths are unequal. One can show that the leading Shapiro delay signal in $k/\omega$ expansion, Eq.~\eqref{shapiro_leading}, cancels in the 1.5 generation TDI variables even when $L_{ij}\neq L$.} The Shapiro signal in TDI variables therefore arises at next-to-leading order in the $k/\omega$ expansion:
\begin{equation}
    \label{shapiro_fast_nlo}
    \tilde s_{ij}(f) = 
    -i\omega_L \int \frac{d^3\boldsymbol{k}}{(2\pi)^3} 
    \frac{\boldsymbol k \cdot \hat {\boldsymbol n}_{ij}}{\omega^2} 
    {\cal R}_{ij}
    \left[ \tilde\Phi(k) +\tilde \Psi(k) \right]e^{i \boldsymbol k \cdot \boldsymbol x_i},
\end{equation}
where, for brevity, we have omitted the arguments of the response function. Therefore, comparing Eqs.~\eqref{eq:one_way_pert_acc} and \eqref{shapiro_fast_nlo}, we see that the Shapiro time delay signal due to coherent fluctuations that would appear in the final TDI variables is of the same order as the acceleration signal. 

Consequently, the dominant ULDM signal is given by
\begin{equation}\label{sij_freq}
    \tilde s_{ij}(f) 
    \approx  \, 
    i\omega_L\!
    \int\frac{d^3\boldsymbol{k}}{(2\pi)^3}
    \frac{\boldsymbol k \cdot \hat{\boldsymbol n}_{ij}}{\omega^2}
    {\cal R}_{ij}
    \tilde{\mathcal{U}}(k)e^{i \boldsymbol k \cdot \boldsymbol x_i}\,,
\end{equation}
where $\tilde{\mathcal{U}}(k)$ is given by
\begin{equation}
    \\
    \tilde{\mathcal{U}}(k)=
    \begin{cases}
    - \widetilde\Psi (k)
    & \textrm{grav. (fast)}
    \\[4pt]
    + \widetilde\Phi (k)
    & \textrm{grav. (slow)}
    \\[4pt]
    g\,\widetilde{\phi^2} (k)/(2\Mpl^2)
    & \textrm{non-grav. }
    \end{cases}\,.
    \nonumber
\end{equation}   
This approximate expression is derived by noting that the Shapiro delay and acceleration signals are of the same order for the coherent gravitational signal, while the acceleration contribution dominates for all other signals. Note also that $\tilde\Phi(k) \approx \tilde\Psi(k)$ for stochastic gravitational signal. The expression for the non-gravitational interaction is valid both for fast and slow mode fluctuations.

Using the above results, we now quantify the DM signal in terms of its frequency power spectrum. The cross-spectral density of the phase observable of different single-link measurements is defined as
\begin{align}
    \langle \tilde s_{ij}(f) \tilde s_{\ell m}^*(f') \rangle
    = \frac{1}{2}\delta(f-f') S_{ij,\ell m}(f)\,.
    \label{psd_sij}
\end{align} 
From \eqref{sij_freq}, we find the cross-spectral density of the DM signal as
\begin{align}\label{eq:master}
    S_{ij,\ell m}^{\scriptscriptstyle\rm DM}(f)
    \!=\! \frac{2\omega_L^2}{\omega^4}
    \!\!\int\!\!\frac{d^3\boldsymbol{k}}{(2\pi)^3}
    (\boldsymbol k\! \cdot\! \hat{\boldsymbol n}_{ij})
    (\boldsymbol k\! \cdot\! \hat{\boldsymbol n}_{\ell m})
    {\cal R}_{ij} {\cal R}_{\ell m}^*
    P_{\mathcal{U}}(k)
    e^{i{\bm k}\cdot \boldsymbol{L}_{i\ell}}
\end{align}
where $P_{\cal U}(k)$ is a four-dimensional power spectrum of the potential ${\cal U}$, defined via its two-point function, $\langle \tilde {\cal U}(k) \tilde{\cal U}^*(k') \rangle = (2\pi)^4 \delta^{(4)}(k-k') P_{\cal U}(k)$. The potential power spectrum can be further expressed in terms of the power spectrum of the quadratic operators:
\begin{equation}\label{psd_potentials}
     P_{\mathcal{U}}(k) = 
    P_{\phi^2}(k) \times
    \begin{cases}
        \left(\pi G\right)^2 & \textrm{grav. (fast)}
        \\[4pt]
        \left(\pi G\right)^2 \left(4m^2_\phi/|\boldsymbol{k}|^2 \right)^2 & \textrm{grav. (slow)} 
        \\[4pt]
        g^2 /(4\Mpl^4) & \textrm{non-grav.} 
    \end{cases},
\end{equation}
where the non-gravitational relation applies to both fast and slow modes. This expression connects the frequency power spectrum of the observable to the power spectrum of the quadratic operator. An explicit expression for the single-link ULDM signal spectrum, obtained after performing the momentum integral, is given in App.~\ref{app:tdi}.

\subsection{Time Delay Interferometry}\label{subsec:tdi}
The single-link signal derived in Sec.~\ref{subsec:single_link} is completely dominated by laser frequency noise, as well as the noise from the displacements of the optical benches. However, it is possible to construct linear combinations of these single-link observations that suppress these dominant noise sources while preserving ULDM and GW signals, a procedure known as Time-Delay Interferometry (TDI)~\cite{giampieriAlgorithmsUnequalarmMichelson1996, Tinto:1999yr, armstrongTimeDelayInterferometrySpacebased1999}.
In this section, we discuss how the single-link signals derived in the previous section appear in TDI variables typically used for LISA data analysis. We start by reviewing the general mapping between the power spectrum of the single-link signal and the power spectrum of the TDI signal. We then specialize our discussion to the Michelson variables (X, Y, Z) and their quasi-orthogonal combinations (A, E, T).

Consider a single-link data stream $d_{ij}(t)$ that contains both signal and noise. A generic TDI variable is constructed as a linear combination of single-link data streams~\cite{giampieriAlgorithmsUnequalarmMichelson1996, Tinto:1999yr, armstrongTimeDelayInterferometrySpacebased1999, Hartwig:2023pft}:
\begin{equation}
    U(t)=\sum_{ij\in \mathcal{I}} c^U_{ij} d_{ij}(t)\,,
\end{equation}
where $\mathcal{I}=\{12,23,31,21,32,13\}$ denotes all the satellite pairings that can be used to define single-link measurements. 
Here, the coefficient $c_{ij}^U$ is a linear combination of the product of the time-delay operator $D_{ij}$ with constant weight factors; the time delay operator is defined as $D_{ij} = \exp(-L_{ij} \partial_t)$ such that $D_{ij} f(t) = f(t - L_{ij})$.
The frequency power spectrum is defined via the two-point function of TDI variables:
\begin{equation}
    \langle\tilde U(f) \tilde V^*(f')\rangle=
    \frac12 \delta(f-f')S_{UV}(f)\,.
\end{equation}
where $\tilde U(f) = \sum \tilde c_{ij}^U(f) \tilde d_{ij}(f)$. The cross-spectral density $S_{UV}(f)$ is given by a linear combination of cross-spectral densities of single-link measurements:
\begin{equation}\label{eq:S_UV}
    S_{UV}(f)=\sum_{ij,\ell m\in\mathcal{I}} \tilde c_{ij}^U(f)\tilde c_{\ell m}^{V*}(f) \,S_{ij,\ell m}(f)\,.
\end{equation}

One of the simplest sets of TDI variables is the Michelson-like variables (X, Y, Z). In the following, we focus on these TDI variables and derive the signal power spectrum. The Michelson X variable is defined as
\begin{align}
{\rm X} &= ( s_{13} + D_{13} s_{31} + D_{13} D_{31} s_{12} + D_{13} D_{31} D_{12} s_{21} )
\nonumber\\
& - ( s_{12} + D_{12} s_{21} + D_{12} D_{21} s_{13} + D_{12} D_{21} D_{13} s_{31} )\,,
\end{align}
while Y and Z can be obtained by cyclic permutation of spacecraft indices, $1\to 2 \to 3 \to 1$. We refer the reader to Fig.~3.6 of Ref.~\cite{Otto:2015erp} for a graphical representation. The Michelson-like TDI variable effectively compares the phase of a laser that has propagated along two arms of the constellation but in opposite directions. 

Assuming for simplicity that $L_{ij}=L$, we find the cross-spectral densities of Michelson-like TDI variables as
\begin{align}
    \!\!\! 
    S_{\rm XX}^{\rm \scriptscriptstyle DM}(f)\!&=\!16\sin^2(\omega L)\big[2+\cos^2(\omega L)\big]S_\delta(f)\mathcal{I}_{XX}(f)\,,
    \label{SXX_ULDM}
    \\
    \!\!\!
    S_{\rm XY}^{\rm \scriptscriptstyle DM}(f)\!&=\!
    - 8\sin^2(\omega L)\big[1+2\cos(\omega L)\big]S_\delta(f)\mathcal{I}_{XY}(f),
    \label{SXY_ULDM}
\end{align}
where we introduce the power spectrum for the single-test mass displacement due to ULDM as
\begin{equation}
    S_\delta(f)=\frac{2\omega_L^2}{3\omega^4}\int\frac{d|{\bm k}|}{2\pi^2}|{\bm k}|^4 
    P_{\mathcal{U}}(f,{\bm k})\,,
\end{equation}
as well as the response integrals
\begin{widetext}
    \begin{align}
        \mathcal{I}_{\rm XX}(f)&=\int_0^\infty dx\,p(x) \left\{1-\frac{3}{2+\cos^2(\omega L)}\left[ \left(\frac{j_1(\rho)}{\rho}+\frac{j_2(\rho)}{2} \right) + 2\cos(\omega L)\left( \frac{j_1(\rho)}{\rho}-j_2(\rho)\right)\right] \right\}\,,
        \\
        \mathcal{I}_{\rm XY}(f)&= \int_0^\infty dx\,p(x) \left\{1-\frac{3\cos(\omega L)}{1+2\cos(\omega L)}\left[ \frac{4j_1(\rho)}{\rho}-j_2(\rho)- \cos(\omega L)\left( \frac{j_1(\rho)}{\rho}+\frac{j_2(\rho)}{2}\right)\right] \right\}\,.
        \label{IXY}
    \end{align}
\end{widetext}
Here, we introduce $\rho\equiv xL\sigma m_\phi$, $x = |\boldsymbol k| / m_\phi \sigma$, and
\begin{equation}
    p(x)\equiv\frac{x^4 P_{\mathcal{U}}(f, x m_\phi \sigma)}{\int dx\,x^4 P_{\mathcal{U}}(f,x m_\phi \sigma)}\,.
\end{equation}
We parameterize the DM signal in this way so that, in the short wavelength limit $m_\phi \sigma L \gg 1$, both response integrals approach unity, and the ULDM signal behaves as extra acceleration noise for each of the LISA test masses. Note also that, in the equal arm-length limit $L_{ij}=L$, we find $S_{\rm XX}(f) = S_{\rm YY}(f) = S_{\rm ZZ}(f)$ and $S_{\rm XY}(f) = S_{\rm YZ} (f)= S_{\rm ZX}(f)$; i.e., the above auto-correlation and the cross-correlation, Eqs.~\eqref{SXX_ULDM}--\eqref{SXY_ULDM}, fully parameterize the DM signal in the Michelson TDI variables. See App.~\ref{app:sig_derivation} for detailed derivations.

An arbitrary set of TDI variables is in general not orthogonal, as is already apparent from the non-vanishing cross-correlation of the Michelson TDI variables \eqref{SXY_ULDM}. It is often convenient to work in the orthogonal basis, in which the two-point function of TDI variables is diagonalized as
\begin{equation}
    \langle\tilde U(f)\tilde V(f')^*\rangle=
    \frac{1}{2} \delta(f-f') \delta_{UV} S_{UU}(f)\,.
\end{equation}
For the Michelson variables, these quasi-orthogonal variables are called AET, and are defined by the following transformation:
\begin{align}
    \left(
    \begin{array}{c}
    {\rm A}
    \\
    {\rm E}
    \\
    {\rm T}
    \end{array}
    \right)
    =
    \left(
    \begin{array}{ccc}
    - \frac{1}{\sqrt{2}} & 0 & \frac{1}{\sqrt{2}}
    \\
    \frac{1}{\sqrt{6}} & - \frac{2}{\sqrt{6}} & \frac{1}{\sqrt{6}}
    \\
    \frac{1}{\sqrt{3}} & \frac{1}{\sqrt{3}} & \frac{1}{\sqrt{3}}
    \end{array}
    \right)
    \left(
    \begin{array}{c}
    {\rm X}
    \\
    {\rm Y}
    \\
    {\rm Z}
    \end{array}
    \right)
\end{align}
This transformation matrix diagonalizes the two-point function of Michelson TDI variables in the equal arm-length limit. The form of the ULDM signal power spectrum in the AET basis is related to that in the XYZ basis as
\begin{align}
\label{eq:S_dm_aa}
    S_{\rm AA}(f)&=S_{\rm EE}(f)=S_{\rm XX}(f)-S_{\rm XY}(f)\,,\\
    S_{\rm TT}(f)&=S_{\rm XX}(f) + 2 S_{\rm XY}(f)\,.
\label{eq:S_dm_tt}
\end{align}
Eqs.~\eqref{eq:S_dm_aa}--\eqref{eq:S_dm_tt} hold not only for the signal power spectrum but also for the noise power spectrum. This is because the noise power spectrum of the Michelson variables is also fully characterized by common diagonal and off-diagonal spectra in the equal arm-length limit. 

\section{Analysis}\label{sec:analysis}
To forecast LISA's sensitivity to ULDM signals, we perform a Bayesian analysis on mock data representative of expected LISA noise levels. In this section, we provide details about the noise assumptions made in our analysis (Sec.~\ref{subsec:noise}), the procedure used to generate mock data (Sec.~\ref{subsec:mock_data}), and the analysis of generated mock data (Sec.~\ref{subsec:data_analysis}). 

\subsection{Noise and Foregrounds}\label{subsec:noise}
We generate and analyze mock data at the level of the TDI variables, assuming that all bright resolvable sources have been subtracted from the data stream. The remaining signals are well described as stochastic, Gaussian, and stationary processes. We model the power spectrum of the TDI variables as the sum of three contributions: instrumental noise, $S_{UV}^{\scriptscriptstyle \rm N}$, astrophysical GW foregrounds, $S_{UV}^{\scriptscriptstyle \rm GW}$, and potential ULDM signals, $S_{UV}^{\scriptscriptstyle \rm DM}$. The resulting cross-spectral density can be written as
\begin{equation}
    S_{UV}(f) = S_{UV}^{\scriptscriptstyle \rm DM}(f) + S_{UV}^{\scriptscriptstyle \rm GW}(f) +S_{UV}^{\scriptscriptstyle \rm N}(f)\,.
\end{equation}
For our analysis, we use the quasi-orthogonal AET variables introduced in the previous section.
The form of $S_{UV}^{\scriptscriptstyle \rm DM}$ for the AET channels was already given in Eq.~\eqref{eq:S_dm_aa} and Eq.~\eqref{eq:S_dm_tt}. In the rest of this section, we provide expressions for the instrumental and astrophysical GW foreground contributions. 

The instrumental noise is given by the sum of test mass (TM) acceleration noise and optical metrology system (OMS) noise.  Their contribution to the TDI noise power spectrum is given by~\cite{Flauger:2020qyi}
\begin{widetext} 
\begin{align}
    S_{\mathrm{AA}}^{\scriptscriptstyle N}(f) &=     8\sin^2\!\left(2\pi f L\right)\left\{4\left[1+\cos\!\left(2\pi f L\right)+\cos^2\!\left(2\pi f L\right)\right]S^{\scriptscriptstyle\rm TM}(f) +  \left[2+\cos\!\left(2\pi f L\right)\right]S^{\scriptscriptstyle\mathrm{OMS}}(f)\right\},\\     S_{\mathrm{TT}}^{\scriptscriptstyle N}(f) &= 16\sin^2\!\left(2\pi f L\right)\left\{2\left[1-\cos\!\left(2\pi f L\right)\right]^2 S^{\scriptscriptstyle\rm TM}(f) + \left[1-\cos\!\left(2\pi f L\right)\right]S^{\scriptscriptstyle\mathrm{OMS}}(f)\right\}\,.
\end{align} \end{widetext}
In the above expressions, we have introduced the single-link PSD for the TM and OMS noise as~\cite{Robson:2018ifk, Flauger:2020qyi, Babak:2021mhe}
\begin{align}
S^{\scriptscriptstyle \rm TM} 
&\!=\! \frac{A^2\omega_L^2}{\left({2\pi f}\right)^4} \frac{\rm fm^2}{\rm s^4 Hz}\Bigg[\!1+\left(\frac{0.4\,{\rm mHz}}{f}\right)^2\!\Bigg]\!\Bigg[\!1+\left(\frac{f}{8\,{\rm mHz}}\right)^4\!\Bigg]
\label{TM_noise}
\\     
S^{\scriptscriptstyle \rm OMS} 
&= P^2\omega_L^2\frac{\rm pm^2}{\rm Hz} \Bigg[1+\left(\frac{2\,{\rm mHz}}{f}\right)^4\Bigg]
\,.
\label{OMS_noise}
\end{align}
Here, $A$ and $P$ are unknown parameters. By the LISA mission specifications, these parameters are expected to satisfy $A=3$ and $P=15$ within a $\pm20\%$ tolerance~\cite{Adams:2013qma}. In above expressions, optical metrology and acceleration noise in each spacecraft are assumed to be the same and given by Eq.~\eqref{TM_noise}--\eqref{OMS_noise}. The noise spectra are also written in the phase unit ${\rm rad^2/ Hz}$, which explains the difference in the expressions of the noise power spectra compared to other works.

To quantify the astrophysical GW foreground, we first parameterize the spectrum as
\begin{equation}     
    S_{UU}^{\scriptscriptstyle \rm GW}(f) =
    16 (\omega_L L)^2 \sin^2(2\pi f L)\mathcal{R}_{U}(f) S_h(f)
\end{equation}
where $U$ = \{A, E, T\}, $S_h(f)$ is strain power spectrum, and $\mathcal{R}_U(f)$ is the response function, given by~\cite{Robson:2018ifk, Flauger:2020qyi} 
\begin{align}
    \mathcal{R}_{\rm A}(f) &=\mathcal{R}_{\rm E}(f) \simeq \frac{9}{20}\frac{1}{1+0.7(2\pi fL)^2}\,,
    \\
    \mathcal{R}_{\rm T}(f) &\simeq \frac{9}{20}\frac{(2\pi fL)^6}{1.8\times 10^3+0.7(2\pi fL)^8}\, .
\end{align}
At frequencies $f\lesssim 1/L$, the GW response of the T-channel is suppressed by a factor $(fL)^6$ compared to the response of the A and E channels. Therefore, in the rest of this work, we treat the T-channel as a null channel that we use to estimate the noise properties of the detector. The strain power spectrum can be written as the fractional energy density per logarithmic interval, $\Omega_{\rm GW}(f) = (4\pi^2 f^3 / 3 H_0^2) S_h(f)$.

\begin{table}[t]
\bgroup
\renewcommand{\arraystretch}{1.5}
\setlength\tabcolsep{4pt}
{\footnotesize
	\begin{tabular}{lll}
		\toprule
		\textbf{Par.}	                           &   \textbf{Description}                 &   \textbf{Injected value}   \\ \hline
		$\alpha_{\rm gal}$                         &   galactic foreground amplitude        &   -7.84   \\
		$a_1$                                      &   $f_1$ slope parameter                &   -0.15  \\
		$b_1$                                      &   $f_1$ intercept parameter            &   -2.72  \\
		$a_k$                                      &   $f_{\rm knee}$ slope parameter       &   -0.37   \\
		$b_k$                                      &   $f_{\rm knee}$ intercept parameter   &   -2.49   \\
		$f_2\,[{\rm mHz}]$                         &   knee transition width                & $0.67$   \\
		$\nu$                                      &   high-frequency cutoff exponent       &   1.56  \\
		$\alpha_{\rm EG}$                          &   extragalactic foreground amplitude   &   -12.38 \\
        $f_*\,[{\rm mHz}]$                         &   extragalactic reference frequency   &   1.00  \\ \midrule
        $A$                                        &   acceleration noise  amplitude        &   3   \\
        $P$                                        &   OMS noise  amplitude                 &   15   \\
        
        \bottomrule
	\end{tabular}
}
\egroup
\caption{Values of the noise and astrophysical parameters used to  generate the mock data.}
\label{tab:noise_params}
\end{table} 

The astrophysical foreground consists of unresolved galactic and extragalactic sources: the galactic component is expected to be dominated by white-dwarf binaries, whose spectrum can be parameterized as above~\cite{Cornish:2017vip, Robson:2018ifk, Schmitz:2020rag, Blanco-Pillado:2024aca}, while the extragalactic component arises from unresolved neutron stars and stellar-origin black holes. Their GW energy densities are parameterized as:
\begin{align}
\Omega_{\rm GW}(f)
= \Omega_{\rm GW}^{\rm gal}(f) + \Omega_{\rm GW}^{\rm ex}(f)
\end{align}
where each contribution is given by~\cite{Cornish:2017vip, Robson:2018ifk, Schmitz:2020rag, Blanco-Pillado:2024aca, LIGOScientific:2019vic, Sesana:2016ljz,Kume:2024xvh}
\begin{align}
h^2\Omega_{\rm GW}^{\rm gal}(f) 
&= \frac{10^{\alpha_{\rm gal}}}{2} \!\left(\frac{f}{\rm Hz}\right)^{\!\frac{2}{3}} \!\left[1 + \tanh\frac{f_{\rm knee} - f}{f_2}\right]e^{-(f/f_1)^{\nu}}\,, 
\\
h^2 \Omega_{\scriptscriptstyle \rm GW}^{\rm ex}(f) 
&= 10^{\alpha_{\rm EG}}\left(\frac{f}{f_*}\right)^{2/3}\,.
\end{align}

Here, we also introduce
\begin{align}
\log_{10}\left(\frac{f_1}{\rm Hz}\right) &= a_1 \log_{10}\left(\frac{T_{\rm obs}}{\rm yr}\right) + b_1\,,\\     \log_{10}\left(\frac{f_{\rm knee}}{\rm Hz}\right) &= a_k \log_{10}\left(\frac{T_{\rm obs}}{\rm yr}\right) + b_k\,,
\end{align}
where $T_{\rm obs}$ is the total observing time. When generating the mock data, we set the parameters entering the above equations to the values reported in Table~\ref{tab:noise_params}. 

\subsection{Mock Data}\label{subsec:mock_data}
To investigate the sensitivity to ULDM signals, we simulate AET variables in the absence of a dark matter signal. Specifically, we generate realizations of the AET variables according to the noise power spectrum described in the previous section, estimate their power spectral densities, and use them to derive projected constraints on ULDM.
To this end, we assume a LISA mission duration of 4 years with a 75\% duty cycle, giving 3 years of effective observation time. The data are divided into equal-size segments of length $T=11.5$ days, yielding a total of $N_s=95$ segments.
In each segment, given a frequency resolution of $\Delta f=1/T\simeq10^{-6}\,{\rm Hz}$ and a frequency coverage going from $f_{\rm min}=3\times 10^{-5}\,{\rm Hz}$ to $f_{\rm max}=5\times 10^{-1}\,{\rm Hz}$, we can resolve approximately $5\times10^5$ frequency modes for any given TDI variable.

We generate mock AET variables in the frequency domain by randomly drawing their real and imaginary parts from a normal distribution:
\begin{align}     
    {\rm Re}[\tilde{U}^l(f)] &\sim\mathcal{N}\left(0,\frac{S_{UU}(f)}{4\Delta f}\right)\,,
    \\         
    {\rm Im}[\tilde{U}^l(f)] &\sim\mathcal{N}\left(0,\frac{S_{UU}(f)}{4\Delta f}\right)\,, 
\end{align}
where ${\rm U}= \{\rm A,\, E,\, T\}$, $l$ indexes the data segments, and the power spectrum for the different TDI variables includes the noise and astrophysical GW foreground contributions discussed in Sec.~\ref{subsec:noise} but no ULDM signal. Here ${\cal N}(\mu,\sigma^2)$ represents a normal distribution with mean $\mu$ and standard deviation $\sigma$. For a given realization of $\tilde U^l(f)$, the power spectrum in each segment $l$ is estimated as $D^l_U(f) = 2\Delta f|\tilde U^l(f)|^2$.
The data are then compressed by averaging over data segments to define the average PSD estimators, $\bar D_U(f)$, as
\begin{equation}
    \bar D_U(f)\equiv \frac{1}{N_s}\sum_{l=1}^{N_s}D^l_U(f)\,.
\end{equation}

In contrast to the coherent signal, the stochastic signal exhibits a broadband signal at frequencies $2\pi f < m \sigma^2$, making the frequency resolution of the original data unnecessary. Therefore, for the stochastic search, following Refs.~\cite{Caprini:2019pxz, Flauger:2020qyi}, we bin the modes at frequencies larger than $10^{-3}\,{\rm Hz}$ into 1000 logarithmically-spaced frequency bins, where the $k$-th bin contains $n^{(k)}$ of the initial frequency modes. The new coarse-grained frequency bins, $f_U^{(k)}$, and the averaged power spectrum estimator, $\mathcal{D}^{(k)}_U$, are given as
\begin{align}
    f_U^{(k)}&\equiv
    \sum_{f\in \Omega_k}
    w_{\scriptscriptstyle U}^{(k)}(f) f\,,\\     
    \mathcal{D}^{(k)}_U&\equiv 
    \sum_{f\in \Omega_k}
    w_{\scriptscriptstyle U}^{(k)}(f) \bar D_U(f)\,, \label{eq:corse_data} 
\end{align}
where the sum $f\in \Omega_k$ runs over the set of frequencies, $\Omega_k$, that belong to the $k$-th bin, and the weights, $w_U^{(k)}(f)$, are given by
\begin{equation}     
w_U^{(k)}(f)=\frac{\sigma_U^{-2}(f)}{\sum_{f\in \Omega_k}\sigma_U^{-2}(f)}\,, 
\end{equation}
Here we choose $\sigma_U^2(f) = [S^{\scriptscriptstyle \rm N}_U(f)]^2/N_s$. Note that, since the weights $w_{\scriptscriptstyle U}^{(k)}$ differ across TDI channels, the corresponding frequency bins, $f_U^{(k)}$ are also channel-dependent. 

\begin{table}[t!]
\bgroup
\renewcommand{\arraystretch}{1.5}
\setlength\tabcolsep{4.pt}
{\footnotesize
	\begin{tabular}{lll}
		\toprule
		\multicolumn{1}{c}{\textbf{Parameter}}	   &   \multicolumn{1}{c}{\textbf{Prior}}                &   \multicolumn{1}{c}{\textbf{Comments}}   \\ 
		\midrule
		$\alpha_{\rm gal}$                         &   $\mathcal{N}(-7.84, 0.21^2)$          &     \\
        $a_1$                                      &   $-0.15$                              & assumed known \\
		$b_1$                                      &   $-2.72$                             & assumed known \\
		$a_k$                                      &   $-0.37$                             & assumed known \\
		$b_k$                                      &   $-2.49$                             & assumed known \\
		$f_2\,[{\rm mHz}]$                          &   $0.67$                & assumed known \\
		$\nu$                                      &   $1.56$                              & assumed known \\
		$\alpha_{\rm EG}$                   &   $\mathcal{N}(-12.38, 0.17^2)$         &     \\
        $A$                                        &   $\mathcal{N}(3, 0.6^2)$               & for T-channel noise run    \\
        $P$                                        &   $\mathcal{N}(15, 3^2)$                & for T-channel noise run    \\
        \midrule
		$\log_{10}( m_\phi/{\rm eV})$                              & ${\rm Uniform}[-14.5, -9]$       &  for stochastic signal   \\
		$\log_{10}( m_\phi/{\rm eV})$                              &  ${\rm Uniform}[-19.2, -15]$       &  for coherent signal   \\
		$\log_{10} g$                              &   ${\rm Uniform}[6, 18]$         & for stochastic signal   \\
        $\log_{10} g$                              &   ${\rm Uniform}[3, 16]$         & for coherent signal   \\
        $\log_{10}(\rho/\rho_0)$                              &   ${\rm Uniform}[0, 16]$         & for stochastic signal   \\
        \bottomrule
	\end{tabular}
}
\egroup
\caption{Model parameters and prior distributions used in the Bayesian analysis. The priors for the astrophysical parameters are chosen following the same conventions adopted in Ref.~\cite{Blanco-Pillado:2024aca}. The constraints derived from the coherent signal in the gravitational coupling case are derived by replacing $g\xrightarrow{}\rho/(2\bar{\rho})$, as the spectral shapes are identical up to this amplitude reparametrization.}
\label{tab:priors}
\end{table}

\subsection{Bayesian Analysis}\label{subsec:data_analysis} 
To forecast LISA's sensitivity to the ULDM signals discussed in Sec.~\ref{sec:uldm_signal}, we perform a Bayesian analysis on the mock data generated following the procedure outlined in the previous section.

We approximate the likelihood function as~\cite{WMAP:2003pyh}
\begin{align}
    \ln \mathcal{L}({\bm\eta}|\boldsymbol{{\mathcal{D}}})=
    \frac{1}{3}\ln\mathcal{L}_G({\bm\eta}| {\mathcal{\bm D}})
    +\frac{2}{3}\ln\mathcal{L}_{LN}({\bm\eta}|{\mathcal{\bm D}})\,.
\end{align}
where each term is given by
\begin{align}
    \ln\mathcal{L}_G({\bm\eta}|{\boldsymbol{\mathcal{D}}})
    &\approx
    -\frac{N_s}{2}
    \sum_U\sum_k
    n^{(k)}
    \left[\frac{S_{UU}(f_U^{\rm (k)};{\bm\eta})-\mathcal{D}^{(k)}_U}{S_{UU}(f_U^{\rm (k)};{\bm\eta})}\right]^2\,,
    \\
    \ln\mathcal{L}_{LN}({\bm\eta}|{\mathcal{\bm D}})
    &=
    - \frac{N_s}{2}\sum_U\sum_kn^{(k)}
    \ln^2\left[\frac{S_{UU}(f_U^{\rm (k)};{\bm\eta})}{\mathcal{D}^{(k)}_U} \right]\,.
\end{align}
Here, ${\bm \eta}$ denotes the model parameters, summarized in Table~\ref{tab:priors} together with the corresponding prior distributions used in the analysis. The first term is the Gaussian approximation, and the second term is the log-normal correction to the first to prevent systematic bias in parameter estimation~\cite{Bond:1998qg, Sievers:2002tq, WMAP:2003pyh, Hamimeche:2008ai, Franciolini:2025leq}. The relative numerical coefficient is chosen such that the above approximate likelihood function matches the full one up to ${\cal O}(\delta^3)$ where $\delta = (D-S)/S$.

\begin{figure}[t]
\centering
\includegraphics[width=0.48\textwidth]{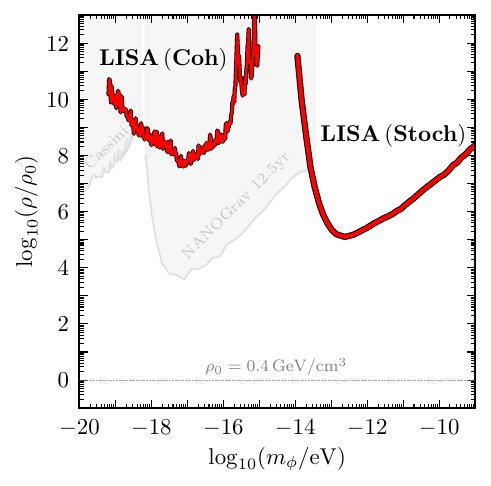}
\caption{Projected LISA sensitivities to a gravitationally coupled ULDM field (red lines). The curve on the left side is for the coherent signal search, while the one on the right side is for the stochastic signal search. For comparison, we show the NANOGrav 12.5-year constraint~\cite{Kim:2023kyy} and Cassini constraint~\cite{Blas:2016ddr, Armstrong:2003ay} as the light-gray shaded region. We also include the benchmark line implied by the local dark-matter density, $\rho_0 = 0.4\,\mathrm{GeV}/\mathrm{cm}^3$.}
\label{fig:main_density}
\end{figure}

To derive the posterior distributions for the parameters of the ULDM signal, we proceed in two steps. First, we derive posterior distributions for the noise parameters $A$ and $P$ by performing MCMC runs using the above likelihoods on T channel data alone. This auxiliary step uses the T channel to calibrate the instrumental noise since gravitational wave and dark matter signals are suppressed in this channel.

Then, we perform a second MCMC run in which we use A and E channel data to derive posterior distributions for all model parameters, including those associated with the ULDM signal. In this second run, the posterior distributions for $A$ and $P$ obtained from the T channel analysis are used as priors for the noise parameters.

\begin{figure*}[t]
\centering
\includegraphics[width=\textwidth]{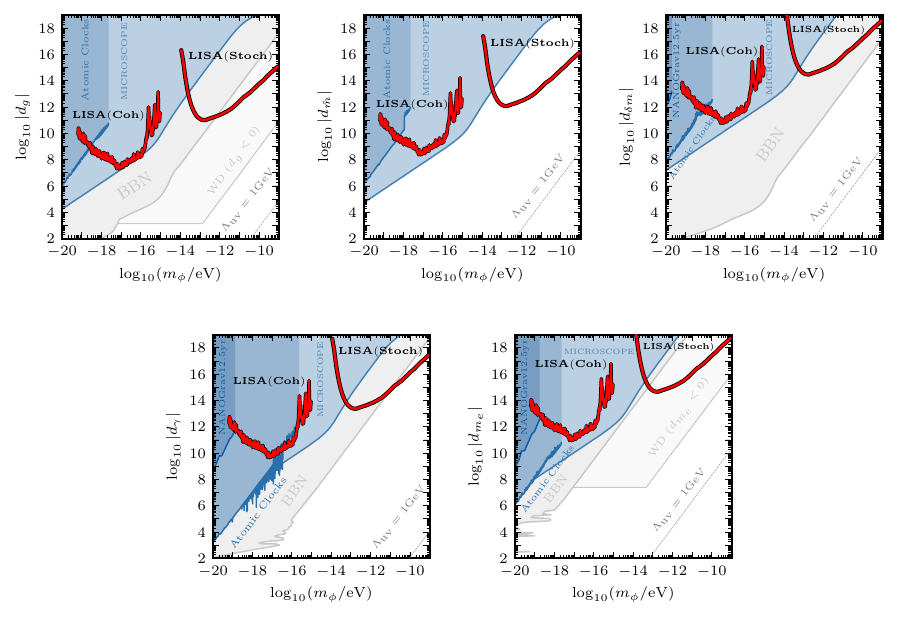}
\caption{Projected LISA sensitivities to quadratically-coupled dilaton-like ULDM with couplings $d_i$ (red lines). Other constraints --- from MICROSCOPE~\cite{Hees:2018fpg,MICROSCOPE:2019jix,MICROSCOPE:2022doy,Bouley:2026frx,Brzeminski:2026pgz}, BBN~\cite{Stadnik:2015kia,Sibiryakov:2020eir,Bouley:2022eer,Ghosh:2025pbn}, atomic clock comparison tests~\cite{Hees:2016gop, Kennedy:2020bac, BACON:2020ubh, Oswald:2021vtc, Sherrill:2023zah, Filzinger:2023zrs}, and the NANOGrav 12.5-year data~\cite{Gan:2025icr} --- are also shown. We have also included limits from structural changes in white dwarfs (WDs) induced by quadratic couplings to electrons and gluons~\cite{Bartnick:2025lbg,Bartnick:2025lbv}. Note that the WD constraints apply to negative (attractive) couplings, whereas the other constraints are either insensitive to the sign of $d_i$ or apply to positive (repulsive) couplings. Gray dashed lines are obtained by requiring that quantum corrections to the mass of the scalar particle are  smaller than its own mass when the cutoff scale is $\Lambda_{\rm UV}=1\GeV$.}
\label{fig:main_dilaton}
\end{figure*}

\section{Results}\label{sec:results}
\subsection{ULDM density constraints}
We begin by presenting the results of the ULDM search through gravitational interactions. In Fig.~\ref{fig:main_density}, we show the projected LISA sensitivity to the ULDM density within the solar system. The constraints in the mass range $10^{-19}\,{\rm eV}$–$10^{-15}\,{\rm eV}$ derive from the search for coherent signals, while the constraints in the region $10^{-14}\,{\rm eV}$–$10^{-9}\,{\rm eV}$ derive from the search for stochastic ULDM signals. Our coherent-signal result generally agrees with the projections presented in previous works~\cite{Aoki:2016kwl, Kim:2023pkx, Yu:2024enm}, where the sensitivity was obtained by comparing the signal and noise power spectral densities instead of performing a Bayesian analysis on mock data. The result for the stochastic signal shows that even a single LISA constellation can probe the solar-system dark matter density at a level comparable to that of Ref.~\cite{Kim:2023pkx}, which assumed the cross-correlation of multiple LISA-like constellations.

It is important to emphasize that LISA observations will be able to constrain the DM abundance in the solar system, a quantity that lacks direct experimental measurement. Current estimates of the local DM abundance are inferred from large-scale properties of the Milky Way, probing volumes of $\mathcal{O}(10^6\,{\rm pc}^3)$ or larger (for reviews, see Refs.~\cite{Read:2014qva, deSalas:2020hbh}). Consequently, the canonical value $\rho_0 \sim 0.4\,{\rm GeV/cm}^3$ represents a large-scale average and does not preclude the existence of significant DM overdensities on much smaller scales.
To date, the most stringent constraints on the DM abundance in the vicinity of the solar system, $\rho_\oplus$, come from PTA searches for ULDM~\cite{Kim:2023kyy}. 
An analysis of solar system ephemerides also constrains the density of a static, spherically symmetric dark matter distribution centered on the Sun~\cite{Pitjev:2013sfa}. In such a scenario, the dark matter density is constrained to be smaller than $\rho_\oplus/\rho_0 \lesssim 2 \times 10^5$ at Earth's orbit and $\rho_\oplus/\rho_0 \lesssim 2 \times 10^4$ at the Mars/Saturn orbits. These limits do not directly apply to our case, although they might provide a meaningful constraint for other ULDM scenarios, e.g. ULDM captured by the Sun via self-interactions~\cite{Budker:2023sex, Banerjee:2019epw, Banerjee:2019xuy}. Analyses of the interplanetary range measurements also provide a way to probe ultralight dark matter gravitationally for the mass range around $10^{-15}\eV$ with sensitivity $\rho / \rho_0 \lesssim 10^5$~\cite{Frerick:2026kzq}, while Cassini Doppler tracking provides a limit at the lower mass range of the parameter space~\cite{Blas:2016ddr, Armstrong:2003ay}.

\subsection{Scalar couplings constraints}
Next, we report the constraints on ULDM direct coupling to the SM. We first consider a dilaton-like ULDM whose couplings to the SM are given by
\begin{align}\label{eq:lagrangian}
    {\cal L}
    = 
    \frac{\phi^2}{2\Mpl^2}
    \bigg[
    \frac{d_\gamma}{4e^2}F_{\mu\nu}F^{\mu\nu} 
    - \frac{d_g\beta_3}{2 g_3} G_{\mu\nu}^AG_A^{\mu\nu}
    \nonumber\\
    - \sum_\psi (d_{m_\psi}+\gamma_{m_\psi} d_g) m_\psi \bar \psi \psi
    \bigg]\,,
\end{align}
where $\beta_3$ denotes the beta-function coefficient of the strong sector, $\gamma_{m_\psi}$ are the anomalous dimensions of the light quarks, and $(d_g, d_\gamma, d_{m_\psi})$ are dimensionless coupling constants. The sum over $\psi$ runs over the electron and the up and down quarks. 

For this model, the coupling parameter that enters Eq.~\eqref{eq:acc} can be written as
\begin{align}
g = \frac{\partial \ln M}{\partial (\phi^2/2\Mpl^2)}
= \boldsymbol d \cdot \boldsymbol Q
\end{align}
where $\boldsymbol d$ is the basis of the quadratic couplings introduced in Eq.~\eqref{eq:lagrangian}:
\begin{equation}
\label{eq:vec_d}
    \bm{d}
    =
    (
        d_g, \,
        d_\gamma, \,
        d_{\hat m} - d_g, \,
        d_{\delta m} - d_g, \,
        d_{m_e} - d_g
        )\,,
\end{equation}
and
\begin{align}
    d_{\hat m} &= 
    \frac{d_{m_d} m_d + d_{m_u} m_u}{m_u + m_d}\,,
    \\
    d_{\delta m} &= 
    \frac{d_{m_d} m_d - d_{m_u} m_u}{m_d - m_u}\,,
\end{align}
with $m_u$ and $m_d$ being the up- and down-quark masses, and $\hat{m} = (m_u + m_d)/2$ and $\delta m = m_d - m_u$ denoting their symmetric and antisymmetric combinations, respectively. For LISA test masses, which are composed of a gold–platinum alloy, the dilaton charge vector, $\bm{Q}$, is given by~\cite{Damour:2010rm}
\begin{equation}
\label{eq:dilaton_charge_LISA}
\boldsymbol{Q}  = (1, 4 \times 10^{-3}, 9 \times 10^{-2}, 3 \times 10^{-4}, 2 \times 10^{-4})\,,
\end{equation}
where the entries are listed in the same order as in \cref{eq:vec_d}.

The projected LISA sensitivity to each of the dilaton couplings is reported in Fig.~\ref{fig:main_dilaton}, together with other existing constraints from MICROSCOPE~\cite{Hees:2018fpg,MICROSCOPE:2019jix,MICROSCOPE:2022doy,Bouley:2026frx,Brzeminski:2026pgz}, atomic clock comparison tests~\cite{Hees:2016gop, Kennedy:2020bac, BACON:2020ubh, Oswald:2021vtc, Sherrill:2023zah, Filzinger:2023zrs}, big bang nucleosynthesis~(BBN)~\cite{Stadnik:2015kia,Sibiryakov:2020eir,Bouley:2022eer,Ghosh:2025pbn,Aghaie:2026bqq}, and PTAs~\cite{Gan:2025icr}. As shown in Fig.~\ref{fig:main_dilaton}, the low-mass constraints derive from the coherent ULDM signal, while those at higher masses derive from the stochastic signal. All results shown in the figure are obtained by assuming that only one dilaton coupling is nonzero. The gravitational contribution is subdominant in the parameter space shown and is therefore not included.

\subsection{Axion couplings constraints}
Finally, we consider QCD axions, whose coupling to SM particles is described by the Lagrangian
\begin{equation}
{\cal L} = \frac{g^2_3}{32\pi^2} \frac{\phi}{\fphi} G_{\mu\nu}^a \widetilde G^{a\mu\nu}\,,
\end{equation}
where $\widetilde G_{\mu\nu}^a$ is the dual of $G_{\mu\nu}^a$, and $\fphi$ is the axion decay constant. Below the QCD scale, the strong sector confines, explicitly breaking the underlying axion shift symmetry. Consequently, the axion develops quadratic couplings to mesons and hadrons~\cite{Kim:2022ype}, as well as to the photon and the electron fields via loop corrections~\cite{Beadle:2023flm, Kim:2023pvt, Gan:2025nlu, Bai:2025yxm}. For the case of LISA, the dominant signal is produced by the quadratic coupling to nucleons, described by the effective Lagrangian
\begin{equation}\label{axion_quad}
    {\cal L} = 
    - C_N \frac{\phi^2}{2 \fphi^2} m_N \bar N N , 
\end{equation}
where \(N=(p,n)\) is the nucleon field and \(C_N \sim -10^{-2}\)~\cite{Ubaldi:2008nf,Kim:2022ype}. As a consequence of this coupling, oscillations of the axion field will induce oscillations of the nucleon masses:
\begin{equation}
    \frac{\delta m_N}{m_N}=C_N \frac{\phi^2}{2f_\phi^2}\,.
\end{equation}
For the light QCD axion, the value of the effective coupling entering in Eq.~\eqref{eq:acc} is then given by $g=C_N M_{\rm Pl}^2/f_\phi^2$~\cite{Ubaldi:2008nf,Kim:2022ype}. 

\begin{figure}[t]
\centering
\includegraphics[width=\linewidth]{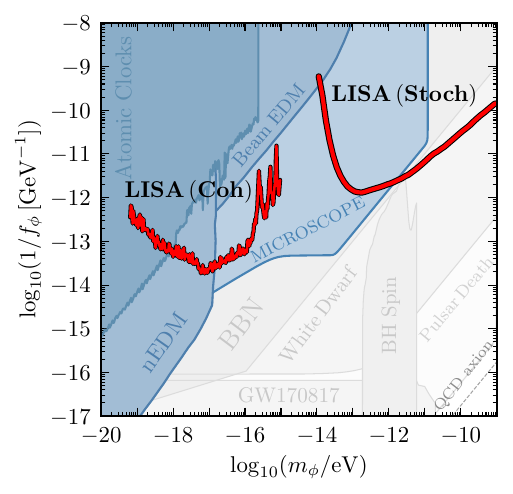}
\caption{Projected LISA sensitivity to a light QCD axion (red lines). Other limits, e.g. those from MICROSCOPE~\cite{Gue:2025nxq,Gan:2025icr}, atomic clock comparison tests~\cite{Hees:2016gop, Kennedy:2020bac, BACON:2020ubh, Sherrill:2023zah, Filzinger:2023zrs, Oswald:2021vtc}, oscillating nEDM~\cite{Abel:2017rtm}, beam EDM~\cite{Schulthess:2022pbp}, BH spin-down~\cite{Arvanitaki:2014wva, Baryakhtar:2020gao, Hoof:2024quk, Witte:2024drg}, GW170817~\cite{Zhang:2021mks}, pulsar death~\cite{Witte:2025ilt}, BBN~\cite{Blum:2014vsa}, and white dwarf mass-radius relation~\cite{Balkin:2022qer}, are also shown.}
\label{fig:main_axion}
\end{figure}

In Fig.~\ref{fig:main_axion}, we report the upper limits on the inverse axion decay constant obtained in our analysis, together with other existing constraints. As shown in the figure, the sensitivity of LISA does not reach the minimal QCD axion line, represented by the gray dashed line in the bottom-right corner of the plot. In other words, LISA will probe either a fine-tuned parameter space of the QCD axion or models where the axion mass is parametrically suppressed compared to that of the QCD axion due to additional particle contents and symmetries in the model construction~\cite{Hook:2018jle, DiLuzio:2021pxd}.

\section{Discussion}\label{sec:discussion}

\subsection{Matter Effects}\label{subsec:matter_effects}
To derive the ULDM signals discussed in the previous sections, we used the plane wave expansion of the field \eqref{plane_wave_expansion} with the boundary condition fixed by the dark matter density in the solar system and the normal velocity distribution. However, in the presence of quadratic couplings, the field profile could be significantly altered near test masses due to the test masses themselves or the surrounding environment. This effect is referred to as screening or matter effects~\cite{Hees:2018fpg, Banerjee:2022sqg, Day:2023mkb, VanTilburg:2024xib, Banerjee:2025dlo, delCastillo:2025rbr, Gan:2025nlu, Burrage:2025grx, Grossman:2025jub, Brzeminski:2026rox,Bouley:2026frx,Brzeminski:2026pgz}, which could suppress the sensitivity of a given experiment. In this section, we argue why, unlike many terrestrial and astrophysical probes, LISA is immune to this screening effect.

Matter effects arise because interactions with ordinary matter induce an additional effective mass for the ULDM field, thereby modifying its profile. 
Specifically, the in-medium ULDM mass squared is given by
\begin{equation}
\label{eq:msq_tot_eq_min}
m_\text{tot}^2 ({\boldsymbol x}) = m_\phi^2 + \Delta m^2 ({\boldsymbol x}), 
\end{equation}
where $\Delta m^2$ is a finite-density correction to the mass, given by
\begin{equation}\label{eq:Del_mMSq}     
    \Delta m^2 ({\boldsymbol x}) = \frac{g \,\rho_{\rm SM}({\boldsymbol x})}{\Mpl^2}\,.
\end{equation}
Here $\rho_{\rm SM}$ indicates the density of ordinary matter to which the ULDM couples. For the following discussion, we focus on the repulsive case $\Delta m^2 > 0$ to illustrate the screening effect~\cite{Hees:2018fpg,Banerjee:2022sqg,Day:2023mkb,VanTilburg:2024xib,Gan:2025nlu,Brzeminski:2026rox,Bouley:2026frx,Brzeminski:2026pgz}. The discussion applies equally to the attractive case $\Delta m^2 < 0$ away from resonance, with the caveat that the attractive potential induces narrow resonance features at $|\Delta m| \, R = \left(n + 1/2\right)\pi$~\cite{Banerjee:2025dlo,delCastillo:2025rbr}.

The mass correction in Eq.~\eqref{eq:Del_mMSq} plays the role of an effective potential for the propagation of ULDM in a medium. Two limits can be considered to understand the behavior of the ULDM profile in the presence of such mass corrections. Consider a spherical object of size $R$ with a density $\rho_{\rm SM}(\boldsymbol x)$. In the limit $\Delta m R\to0$, the potential induced by the quadratic coupling becomes negligible, and ULDM propagates as in vacuum. By contrast, when $\Delta m R\gg1$, the object behaves as a hard sphere, reflecting the ULDM field outward and strongly suppressing the dark matter profile near and within the object. As a specific example, let us consider the planet Earth, $R_\oplus \simeq 6 \times 10^3\,{\rm km}$ and $\rho_\oplus \simeq 5 \, \text{g}/\text{cm}^3$~\cite{moritz2000geodetic,IAUInter-DivisionA-GWorkingGrouponNominalUnitsforStellarPlanetaryAstronomy:2015fjh}. In this case, screening effects become relevant for couplings of order $|g| \gtrsim 10^9$~\cite{Gan:2025icr,Gan:2025nlu,Bouley:2026frx,Brzeminski:2026pgz}; the dark matter field is suppressed near and within Earth, and thus the dark matter signal does not necessarily increase as the coupling constant increases above this value.

The dark matter signal in LISA is free from screening effects. Since LISA operates at least a million kilometers away from any planet in the solar system, the dark matter field profile near the test masses is expected to follow that of the unperturbed halo dark matter. Moreover, unlike many astrophysical observations, it employs compact test masses whose size is small enough that screening effects remain negligible. For instance, substituting the test mass density $\rho_{\rm LISA}$ and its size $R_{\rm LISA}$ into Eq.~\eqref{eq:Del_mMSq}, we find the screening effect remains negligible as long as
\begin{equation}
    \label{eq:LISA_unscreen_regime}
    |g| \lesssim 10^{26} \times \left(\frac{5\,{\rm cm}}{R_{\rm LISA}}\right)^2 \left(\frac{20\,\text{g}/\text{cm}^3}{\rho_{\rm LISA}}\right)\,,
\end{equation}
where the values for the LISA test mass size and density are taken from Ref.~\cite{LISA:2024hlh}.
Recasting Eq.~\eqref{eq:LISA_unscreen_regime} in terms of 
the parameter space for the dilaton-like particle and the light QCD axion, we find that the entire parameter space covered in this work lies well below the screening regime. Consequently, the LISA test masses can be treated as point particles, and the plane-wave approximation remains valid for computing the power spectrum.

\subsection{Cross-Correlation}
If multiple space-based detectors are online simultaneously, their data streams could be cross-correlated to help discriminate a ULDM signal from background noise. Indeed, as already shown in Ref.~\cite{Kim:2023pkx}, the gravitational signal produced by an ULDM field -- due to the long-range nature of the gravitational interaction -- retains significant correlations between different satellites even when they are separated by distances larger than the ULDM coherence length.
On the other hand, for direct coupling interactions, these correlations are strongly suppressed when the separation between the two satellite constellations exceeds the typical dark matter wavelength. In this case, cross-correlations between satellites are of little use to discriminate ULDM signals from the noise.

To illustrate this point, let us consider the position fluctuations of two test masses, $\boldsymbol{\delta x}_a$ and $\boldsymbol{\delta x}_b$,  projected along two arbitrary directions, $\hat{\boldsymbol n}_a$ and $\hat{\boldsymbol n}_b$:
\begin{equation}
    \langle 
    \tilde{\delta x}_{n_a}(f)
    \tilde{\delta x}_{n_b}(f')
    \rangle
    = \frac{1}{2} \delta(f-f') S_{ab}(f)\,,
\end{equation}
where $\delta x_{n_{a,b}}(t) \equiv \hat{\boldsymbol n}_{a,b} \cdot \boldsymbol{\delta x}_{a,b}(t)$. The cross power spectral density, $S_{ab}(f)$, can be obtained similarly to Eq.~\eqref{eq:master}:
\begin{equation}
    S_{ab}(f) 
    = 
    \frac{2}{\omega^4} \int \frac{d^3 {\boldsymbol k}}{(2\pi)^3} e^{i \boldsymbol k \cdot \boldsymbol{L}_{ab}} (\boldsymbol k \cdot \hat{\boldsymbol n}_a )
    (\boldsymbol k \cdot \hat{\boldsymbol n}_b ) P_{\mathcal{U}}(k)\,.
\end{equation}
Due to the oscillatory term, $e^{i\boldsymbol k \cdot \boldsymbol L_{ab}}$, the integral receives the dominant contribution when $|\boldsymbol k| < L_{ab}^{-1}$. In other words, for stochastic signals, the spectrum may be estimated as
\begin{align}
    S_{ab}(f) \sim \int^{L_{ab}^{-1}}_{\omega / \sigma} dk \,  k^4 P_{\mathcal{U}}(k)
    \propto
    \begin{cases}
    \int dk / k
    & \textrm{grav.}
    \\
    \int dk \, k^3
    & \textrm{non-grav.}
    \end{cases} . 
\end{align}
From the structure of the phase space integral, we see that the spectrum receives a similar contribution in each decade of Fourier momentum for the case of gravitational interaction, while the non-gravitational interaction exhibits a strong phase space suppression when the typical wavelength becomes shorter than $L_{ab}$. This behavior can be attributed to the long- and short-range nature of gravitational and non-gravitational interactions. 

For the stochastic signal search, LISA is sensitive to dark matter masses above $10^{-13}\eV$. At this scale, the coherence length is already comparable to the LISA arm length,
\begin{equation}
    \ell_c\simeq\frac{1}{m_\phi\sigma}\simeq3\times10^{6}\,{\rm km}\left(\frac{10^{-13}\,{\rm eV}}{m_\phi}\right)\,.
\end{equation}
For most of the mass range shown in our main figures, the dark matter wavelength is therefore shorter than the typical separation between future space-based interferometer missions (e.g., LISA, TianQin and Taiji, and DECIGO), which would operate at distances of a few AU. As a result, inter-detector cross-correlations remain unsuppressed only for gravitational interactions.

\subsection{Other Subleading Signals}\label{subsec:sub}
For an ULDM field directly coupled to the SM, signals other than the acceleration and Shapiro signals discussed above will be present in the LISA data stream. In this section, we review some of these signals and discuss why for quadratically coupled ULDM they are subleading compared to the acceleration signal.

\emph{Test mass size change --} As discussed in Ref.~\cite{Grote:2019uvn}, variations in the value of the fine structure constant and electron mass will induce fluctuations in the value of the atomic Bohr radius, $a_B$, and in turn the size of solids, $s\sim N a_B$, where $N$ is the number of lattice spacings in the solid. Specifically, we have
\begin{equation}
    \frac{\delta s}{s}\approx\frac{\delta a_B}{a_B}=-\frac{\delta \alpha}{\alpha}-\frac{\delta m_e}{m_e}\,.
\end{equation}
However, compared to the acceleration signal, the perturbation to the photon travel time induced by this effect is suppressed by a factor
\begin{equation}
    \frac{\delta L_{\rm size}}{\delta L_{\rm acc}}
    \sim
    \frac{\omega^2s}{m_\phi v_\phi}\sim 10^{-9}\left(\frac{s}{5\,{\rm cm}}\right)\left(\frac{m_\phi}{10^{-17}\,{\rm eV}}\right)\,,
\end{equation}
where, in the last step, we use $\omega = 2m_\phi$ for coherent signals. Thus, the signal from test mass size fluctuations is negligible in the mass range of interest for LISA. The same conclusion holds also for stochastic signals.

\emph{Propagation effects --} Interactions between the photon and the ULDM field modify the photon dispersion relation. By seeking a solution of the type $A_\mu(x) = \eps_\mu e^{-i s(x)}$ to the photon equation of motion, one finds that the dispersion relation takes the form
\begin{equation}
    0 = k^2 + i k \cdot \partial \ln I (x)\,,
\end{equation}
where $I(x) = 1 - d_\gamma \phi^2(x) / 2\Mpl^2$. The complex nature of the dispersion relation indicates a dissipative effect caused by the photon propagation in the ULDM medium. This effect induces amplitude modulations of the electromagnetic field. By solving the dispersion relation perturbatively, we find 
\begin{equation}
    \boldsymbol E(\lambda) \approx 
    \sqrt{\frac{I(\lambda_t)}{I(\lambda)}} \boldsymbol E(\lambda_t)\,.
    \label{e_field_amp}
\end{equation}
It is straightforward to check that the above amplitude modulation is subdominant compared to the phase modulation due to the acceleration effect. In addition, the phasemeter is only indirectly sensitive to the amplitude modulation, which leads to additional suppression of the amplitude modulation in realistic observables. The real part of the dispersion relation is the same as the null condition at this linear order in the coupling constant, and hence, we expect that our computation of coordinate time lapse is justified even in the presence of the photon-DM coupling to this order. 

\section{Conclusion}\label{sec:conclusion}
In this work, we have derived the sensitivity of LISA to ULDM fields quadratically coupled to the Standard Model. The quadratic nature of the coupling gives rise to ULDM signals in two widely separated frequency ranges: one associated with coherent oscillations of the ULDM field at $\omega=2m_\phi$, and another associated with stochastic oscillations at $\omega\lesssim m_\phi\sigma^2$. This distinctive spectral structure allows LISA to probe two separate mass ranges: a low-mass regime spanning $m_\phi=10^{-19}\,{\rm eV}$ to $10^{-15}\,{\rm eV}$ via coherent signals, and a higher-mass regime from $m_\phi=10^{-14}\,{\rm eV}$ to $10^{-9}\,{\rm eV}$ via stochastic signals.

Importantly, LISA offers unique advantages over existing probes. Unlike terrestrial experiments, LISA operates in space, avoiding the environmental screening from Earth that severely suppresses signals in ground-based searches for couplings $|g_\oplus| \gtrsim 10^9$. This enables LISA to place constraints that, particularly in the higher-mass regime, will compete with and in some cases surpass existing bounds from terrestrial experiments, cosmological observations, and pulsar timing arrays. 

\bigskip 

\acknowledgments
We thank M. Pieroni for useful discussions. This work was supported by the Deutsche Forschungsgemeinschaft under Germany's Excellence Strategy - EXC 2121 Quantum Universe - 390833306. AM acknowledges support from a Royal Society University Research Fellowship (URF-R1-251896). AM acknowledges the hospitality of DESY, Hamburg, where a large part of this work was completed. AL acknowledges the support and hospitality of ETH Zurich during the final parts of this project.
\bigskip

\appendix

\section{Signal}
In this appendix, we provide a derivation of the ULDM signal in a single-link detector and the corresponding power spectrum. We also provide a detailed derivation of signal power spectrum of TDI variables as well as the response integral.

\subsection{Signal Derivation}\label{app:sig_derivation}
Consider a single-link detector setup as depicted in Fig.~\ref{fig:worldlines}. The detector consists of two test masses, or two observers, TM$_{i}$ and TM$_{j}$. A laser is transmitted from TM$_{j}$ at coordinate time $t_t$, and is received at TM$_{i}$ at $t$. The transmitted laser is mixed with a local laser at TM$_{i}$, and the phase difference is measured with a phasemeter. The observable is this phase difference,
\begin{align}
    s_{ij}(t) = s_i(t) - s_j(t) 
\end{align}
where $s_{i}(t)$ is the phase of the local and transmitted laser, $E_{i}(t) \propto e^{-i s_{i}(t)}$, with $E_i(t)$ being the amplitude of the electric field of the laser emitted from TM$_i$.

The evolution of the photon phase is governed by the equation of motion, which can be written as
\begin{align}
    M_{\mu\nu} (x^\rho, \partial_\sigma) A^\nu(x) = 0,
\end{align}
where the matrix $M_{\mu\nu}(x^\rho, \partial_\sigma)$ is some differential operator and generally depends on the spacetime coordinates. We seek an approximate solution of type $A_\mu(x) = \eps_\mu e^{-i s(x)}$ where all spatial dependence resides in $s(x)$. The above equation can be approximated as
\begin{align}
    M_{\mu\nu}(x^\rho, k_\sigma) \eps^\nu \approx 0  , 
\end{align}
where $k_\mu = \partial_\mu s$. Here we ignore the spatial variation of $k_\mu$, as we assume that the system is homogeneous on the photon wavelength scale, which is always satisfied in our case. 

Finding a solution to the above eikonal equation can be viewed as an eigenvalue problem. Consider an eigenvector $\eps^\nu_a$ with eigenvalue $D_a(x, k)$~\cite{1962PhRv..126.1899W}:
\begin{align}
    M_{\mu\nu}(x,k) \eps^\nu_a = D_a(x,k)\eps^\nu_a. 
\end{align}
For a nontrivial solution to exist, we require
\begin{align}
    D_a(x,k) = 0. 
\end{align}
The eigenvalue $D_a(x,k)$ can be considered as a Hamiltonian of the system. To illustrate this, let us suppose that a light ray is parameterized by $x^\mu(\lambda)$ and $k_\mu(\lambda)$ with a parameter $\lambda$. The ray is then defined via
\begin{align}
    \frac{dx_{\gamma}^\mu}{d\lambda}
    &= \{ x_{\gamma}^\mu, D_a\} = + \frac{\partial D_a}{\partial k_\mu} , 
    \label{dx}
    \\
    \frac{dk_\mu}{d\lambda}
    &= \{ k_\mu, D_a\} = - \frac{\partial D_a}{\partial x_{\gamma}^\mu} ,
    \label{dk}
\end{align}
where $\{A, B\} = (\partial A / \partial x^\mu)(\partial B / \partial k_\mu) - (\partial A/ \partial k_\mu) (\partial B / \partial x^\mu)$ is a Poisson bracket. The eigenvalue $D_a(x,k)$ can be considered as a generator of evolution along the parameter $\lambda$. The condition $D_a=0$ is preserved along the light path since $\{D_a,D_a\}=0$. In vacuum, $D_a(x,k) = k^2/2$, and the above equation leads to the trivial photon geodesic equation, $dx_{\gamma}^\mu / d\lambda = k^\mu$ and $dk^\mu / d\lambda = 0$. 

The phase of the photon emitted on the $j$ station when it reaches the $i$ station at time $t=x_i^0(\tau)$ can be obtained by integrating $k_\mu=\partial_\mu s$ along the photon worldline:
\begin{align}
    s_j(t) 
    = s_j(t_t) + \int^\lambda_{\lambda_t} d\lambda' \frac{dx_{\gamma}^\mu}{d\lambda'} k_\mu 
    \label{phase_evol}
\end{align}
where $s_j(\lambda_t)$ is the phase at the time of transmission, $t_t=x_j^0(\tau_t)$. The phase difference at reception is therefore
\begin{align}
    s_{ij}(t)
    = s_i(t) - s_j(t_{t})
    - \int_{\lambda_t}^\lambda d\lambda' \, \frac{dx_{\gamma}^\mu}{d\lambda'} k_\mu. 
\end{align}
The last term on the right-hand side vanishes when the photon is propagating in vacuum. 

Suppose now that the laser is continuously emitted at each test mass. The laser frequency measured by the local comoving observer on station $i$ is 
\begin{align}
    \frac{d s_i}{d\tau_i} = u^\mu_i \partial_\mu s_i = \omega_{L,i}+\delta\omega_{L,i}(\tau_i) , 
\end{align}
where $\tau_i$ is the proper time of TM${_i}$, $u^\mu_i = dx^\mu_i /d\tau_i$ is the four-velocity of the observer, $\omega_{L,i}$ is a constant nominal frequency for the laser, and $\delta\omega_{L,i}$ is the laser frequency fluctuation. Using the above equation, the phase $s_i(t)$ can be written as
\begin{align}
    s_i(t) =  
    c_i + \omega_{L,i} \tau
    + \int^{\tau} d\tau_{i} \, \delta\omega_{L,i}(\tau_{i}) , 
\end{align}
where $c_i$ is an integration constant.
Similarly, the phase of the laser on station $j$ at the time of emission is given by
\begin{equation}
    s_j(t_t) =  
    c_j + \omega_{L,j} \tau_t
    + \int^{\tau_t} d\tau_j\, \delta\omega_{L,j}(\tau_j) .
\end{equation}
The phase difference then becomes
\begin{align}
    s_{ij}(t) =& \, \omega_{\rm het} t
    + \omega_{L,j} (t-t_t) 
    - \int^\lambda_{\lambda_t} d\lambda' \frac{dx_{\gamma}^\mu}{d\lambda'} k_\mu
    \nonumber\\
    &
    + \omega_{L,i} \int^{\tau} d\tau_i
    \left( 
        \frac{\delta\omega_{L}}{\omega_{L}} + \Phi - \frac{v^2}{2}
    \right)_i
    \nonumber\\
    &
    - \omega_{L,j} \int^{\tau_t} d \tau_j
    \left( 
        \frac{\delta\omega_{L}}{\omega_{L}} + \Phi - \frac{v^2}{2}
    \right)_j\,, 
\end{align}
where $\omega_{\rm het} =\omega_{L,i}-\omega_{L,j}$, and the subscript in the integrand of the last two line denotes that the quantities -- fractional frequency uncertainty, the potential, and the velocity --  are those of the corresponding observer. For this expression, we have converted the proper time to the coordinate time using
\begin{equation}
    d\tau \approx dt 
    \left(
    1 + \Phi - \frac{v^2}{2}
    \right) . 
\end{equation}
This result reproduces Eq.~\eqref{eq:s_ij_full} in the main text, and is valid up to a linear order in the small parameters $\Phi$ and $v^2/2$.
As argued in the main text, the last two terms take the form of laser phase noise, and thus cancel out in the TDI variables. 

The coordinate time lapse can be computed by integrating the photon geodesic equation perturbatively. Let us expand the photon four-momentum as
\begin{align}
k^\mu(\lambda) = k_0^\mu(\lambda) + \delta k^\mu(\lambda)
\end{align}
where $k_0^\mu(\lambda) = \omega_L (1 , \hat{\boldsymbol n}_{ ij })$ and $\delta k^\mu$ is linear in the metric perturbation. By integrating the geodesic equation, we find
\begin{align}
    k^\mu(\lambda) 
    \approx& 
    k^\mu(\lambda_t) 
    - \int^\lambda_{\lambda_t} d\lambda' \Gamma^\mu_{\;\;\nu\rho} k^\nu_0 k^\rho_0 , 
    \label{k1}
    \\
    x_{ \gamma }^\mu(\lambda)
    \approx& x_{ \gamma }^\mu(\lambda_t) + k^\mu(\lambda_t) ( \lambda - \lambda_t)
    \nonumber\\
    &- \int^\lambda_{\lambda_t} d\lambda' \int^{\lambda'}_{\lambda_t} d\lambda''
    \Gamma^\mu_{\;\;\nu\rho} k_0^\nu k_0^\rho. 
\end{align}
Contracting the second equation with $\omega_{L}^{-1}\eta_{\mu\nu} k_0^\nu$, we find
\begin{align}
    t-t_t 
    =& \hat{\boldsymbol n}_{ ij } \cdot [ \boldsymbol x_\gamma(\lambda) - \boldsymbol x_\gamma(\lambda_t)]
    \nonumber\\
    &
    + \frac{1}{\omega_{L}} \eta_{\mu\nu} k_0^\mu \delta k^\nu(\lambda_t) (\lambda - \lambda_t)
    \nonumber\\
    &
    -\frac{1}{\omega_{L}} \int d\lambda' \int d\lambda'' \, k_{0\mu} \Gamma^\mu_{\;\;\nu\rho} k^\nu_0 k^\rho_0. 
\end{align}
Note that $k_{0\mu} \Gamma^\mu_{\;\;\nu\rho} k^\nu_0 k^\rho_0 \approx \frac{1}{2} \frac{d}{d\lambda} (\delta g_{\mu\nu} k^\mu_0 k^\nu_0)$. With the boundary condition $\boldsymbol x_\gamma(\lambda) = \boldsymbol x_{i}(t)$ and $\boldsymbol x_\gamma(\lambda_t) = \boldsymbol x_{j}(t_t)$ and the null condition $0 = g_{\mu\nu} k^\mu k^\nu = 2 \eta_{\mu\nu} k_0^\mu \delta k^\nu + \delta g_{\mu\nu} k_0^\mu k_0^\nu$, we finally find
\begin{align}
    t - t_t
    = \hat{\boldsymbol n}_{ij} \cdot 
    [ \boldsymbol x_{i}(t) - \boldsymbol x_{j}(t_t)]
    - \omega_{L} \int^\lambda_{\lambda_t} d\lambda'\, (\Phi + \Psi) . 
\end{align}
This reproduces the expression used in the main text~\eqref{eq:time_delay}. 

We conclude this discussion by briefly commenting on the photon propagation in presence of quadratic couplings between ULDM and the photon field, 
\begin{equation}
    {\cal L} = d_\gamma \frac{\phi^2}{8\Mpl^2} F_{\mu\nu} F^{\mu\nu}.
\end{equation}
In the Lorenz gauge, the equation of motion for the gauge field is
\begin{equation}
    0 = (\partial^2 + \partial_\mu \ln I \partial^\mu) A^\nu
    -\partial_\mu \ln I \partial^\nu A^\mu\,,
\end{equation}
where $I(x) = 1 - d_\gamma \phi^2 / 2\Mpl^2$. As before, we seek a solution of the type, $A_\mu(x) = \eps_\mu e^{-i s(x)}$. We find
\begin{equation}
    0 = (- k^2 - i k \cdot \partial \ln I) \eps^\nu  + i \eps \cdot \partial \ln I k^\nu
\end{equation}
Due to the Lorenz gauge condition, $\eps \cdot k = 0$. Taking the inner product between the equation above and $\eps_\nu$, we find
\begin{equation}
    0 = (k^2 + i k \cdot \partial \ln I) \eps^2.
\end{equation}
For a non-trivial solution to exist, we require
\begin{equation}
    D(x,k) = \frac{1}{2} ( k^2 + i k \cdot \partial \ln I) = 0. 
\end{equation}
Using Eqs.\eqref{dx}--\eqref{phase_evol}, we observe that the imaginary part of $D(x,k)$ causes dissipation of the electric field in the dark matter medium, as given in Eq.~\eqref{e_field_amp}.

\subsection{Signal Spectrum}\label{app:tdi}
In this appendix, we provide a detailed computation of the cross-power spectral density $S_{ij,\ell m}(f)$ reported in Eq.~\eqref{eq:master}. We first begin by proving the relation between the power spectrum of potentials $P_{\cal U}(k)$ and that of the quadratic operator $P_{\phi^2}(k)$, i.e. Eq.~\eqref{psd_potentials}. For the non-gravitational interaction, ${\cal U}= g \phi^2/2\Mpl^2$, and hence the proof is trivial. We will focus below on the case of the gravitational interaction. 

In the conformal Newtonian gauge, the potential is related to the density perturbation, $\delta\rho$, via the Poisson equation
\begin{align}
\nabla^2 \Psi \approx 4 \pi G\,\delta \rho
\end{align}
where the density of the field depends quadratically on the underlying dark matter field
\begin{align}
\rho = \frac{1}{2} \dot{\phi}^2 + \frac{1}{2} ( \nabla \phi)^2 + \frac{1}{2} m_\phi^2 \phi^2. 
\end{align}
The density perturbation is defined as $\delta\rho = \rho - \langle \rho \rangle$. 

From the Poisson equation, we find that the power spectrum of the potential and the density perturbation are related as
\begin{align}
P_\Psi(k) = \frac{(4\pi G)^2}{|\boldsymbol k|^4} P_{\delta\rho}(k), 
\label{eq:poisson}
\end{align}
where the power spectrum for the density perturbation is computed in Refs.~\cite{Kim:2023pkx, Kim:2023kyy, Kim:2024xcr}:
\begin{align}
P_{\delta\rho}^{\rm fast}(k) 
&= 
\frac{|\boldsymbol k|^4}{16} 
\frac{2\pi^2\bar\rho^2}{m_\phi^8\sigma^5}
e^{-\frac{\omega-2m_\phi}{m_\phi\sigma^2}}
\sqrt{\frac{\omega-2m_\phi}{m_\phi\sigma^2}-\frac{|\bm{k}|^2}{4 m_\phi^2 \sigma^2}}\,,
\\
P_{\delta \rho}^{\rm slow} (k)
&= 
\frac{2\pi^2 \bar\rho^2}{|\bm{k}| m_\phi^3 \sigma^4}
\exp\bigg[
-\frac{|\bm{k}|^2}{4m_\phi^2\sigma^2} - \frac{\omega^2}{\sigma^2|\bm{k}|^2}
\bigg]\,.
\end{align}
A direct comparison of the power spectrum of the quadratic field operators \eqref{eq:phi_fast} -- \eqref{eq:phi_slow} reveals the following:
\begin{align}
P_{\delta\rho}^{\rm fast}(k) 
&= 
\frac{|\boldsymbol k|^4}{16} 
P_{\phi^2}^{\rm fast}(k) , 
\\
P_{\delta \rho}^{\rm slow} (k)
&= 
m_\phi^4 P_{\phi^2}^{\rm slow}(k) . 
\end{align}
With Eq.~\eqref{eq:poisson}, the relation \eqref{psd_potentials} can be straightforwardly derived.

The fast mode spectrum can be simplified. With the frequency resolution of LISA, $\Delta f \sim 10^{-6}\,{\rm Hz}$, the lineshape of the fast mode signal cannot be resolved in most cases. In this limit, we may coarse-grain the spectrum as
\begin{align}
P^{\rm fast}(f, \boldsymbol k)
= \frac{1}{\Delta f} \int_f^{f+\Delta f} df' \, P^{\rm fast}(f', \boldsymbol k).
\end{align}
The resulting coarse-grained spectrum of the quadratic field operator is
\begin{align}
P_{\phi^2}^{\rm fast}(k)
\approx 
\delta(f-f_m)
\frac{2\pi^2\bar\rho^2}{m_\phi^8\sigma^5}
\frac{1}{4 \sqrt{\pi} \tau} 
e^{ - \frac{|\boldsymbol k|^2}{4m_\phi^2\sigma^2} }
\label{coarse_grained}
\end{align}
where $f_m = m_\phi / \pi$ and $\tau = 1 / m\sigma^2$. A similar operation can be performed for the power spectra of the density fluctuation and potential. 

\subsubsection{Single-Link Detector}
We first derive the cross spectral density in the single-link detector. From Eq.~\eqref{eq:master}, we parameterize the spectral density as
\begin{align}
    S_{ij,\ell m}^{\scriptscriptstyle\rm DM}(f)
    =
    S_{\delta}(f)
    {\cal I}_{ij,\ell m}(f, \boldsymbol L_{ij}, \boldsymbol L_{\ell m} )
    \label{single_link_decomposition}
\end{align}
where the spectrum and the response integral are defined as
\begin{align}
    S_{\delta}(f)
    & \equiv \frac{2 \omega_L^2}{3\omega^4} \int_0^\infty \frac{d|\boldsymbol k|}{2\pi^2} |\boldsymbol k|^4 P_{\cal U}(k) ,
    \label{S_DM}
    \\
    {\cal I}_{ij,\ell m}
    &\equiv 
    3\!\! \int_0^\infty\!\! dx \, p(x) 
    \int \frac{d\Omega}{4\pi}
    (\hat{\boldsymbol k}\cdot \hat{\boldsymbol n}_{ij})
    (\hat{\boldsymbol k}\cdot \hat{\boldsymbol n}_{\ell m})
    {\cal R}_{ij} {\cal R}_{\ell m}^*
    e^{i \boldsymbol k \cdot \boldsymbol L_{i\ell}} . 
\end{align}
Here we have suppressed the arguments of the response functions, introduced the integration variable $x = |\boldsymbol k| / m_\phi \sigma$, and defined $p(x)$ as
\begin{equation}
    p(x)\equiv
    \frac{x^4 P_{\mathcal{U}}(f,x m_\phi \sigma)}{\int_0^\infty dx\,x^4 P_{\mathcal{U}}(f,x m_\phi \sigma)}\,.
\end{equation}
For each of the signals considered in this work, we find
\begin{align}
p_{\rm fast}(x) 
&= \frac{x^4 e^{-x^2/4}}{12\sqrt{\pi}} , 
\\
p_{\rm slow, gr}(x)
&=
\frac{x^{-1} \exp[-x^2/4 - (\omega\tau)^2/x^2]}{\int_0^\infty dx/x \, \exp[-x^2/4 - (\omega\tau)^2/x^2]} ,
\\
p_{\rm slow, ng}(x) 
&= 
\frac{x^{3} \exp[-x^2/4 - (\omega\tau)^2/x^2]}{\int_0^\infty dx \, x^3 \exp[-x^2/4 - (\omega\tau)^2/x^2]} , 
\end{align}
where the subscripts ${\rm gr}$ and ${\rm ng}$ denote the $p(x)$ functions for gravitational and non-gravitational interactions, respectively. In the above equations, we implicitly assume that the power spectrum does not depend on the direction of $\boldsymbol k$, i.e., $P_{\cal U}(k) = P_{\cal U}(f, |\boldsymbol k|)$.

Using the above parametrization, together with the explicit expression for $P_{\cal U}(k)$, we can finally compute the spectrum $S_{\delta}(f)$. For the fast modes, we find
\begin{align}
    S_\delta(f)
    = 
    \frac{2\omega_L^2}{\omega^4}
    \frac{(\pi G \bar\rho)^2}{m_\phi^3}
    \frac{\delta(f-f_m)}{\tau}
    \times
    \begin{cases}
    1
    & \textrm{grav. }
    \\
    4 g^2 
    & \textrm{non-grav.}
    \end{cases}\,,
\end{align}
where we have used the coarse-grained power spectrum~\eqref{coarse_grained}. For the slow mode, we find
\begin{align}
    S_\delta(f)
    &= 
    \frac{2\omega_L^2}{3\omega^4}
    \frac{(4 \pi G \bar\rho)^2}{m_\phi^3 \sigma^4}
    \times
    \begin{cases}
    K_0(\omega\tau)
    & \textrm{grav.}
    \\
    g^2 \sigma^4 (\omega\tau)^2 K_2(\omega\tau) 
    & \textrm{non-grav.} 
    \end{cases}. 
\end{align}

The response integral does not have an analytic expression. However, the angular part of the integral can be computed analytically. Using the identity
\begin{align}
    \label{identity}
    \int \frac{d\Omega_{\boldsymbol k}}{4\pi} 
    e^{i \boldsymbol k \cdot \boldsymbol x} \hat{\boldsymbol k}^i \hat{\boldsymbol k}^j
    = \delta^{ij} \frac{j_1(k x)}{kx} - \hat{\boldsymbol x}^i \hat{\boldsymbol x}^j j_2(kx), 
\end{align}
we find that the angular integral becomes
\begin{widetext}
\begin{align}
    \int \frac{d\Omega}{4\pi}
    (\hat{\boldsymbol k}\cdot \hat{\boldsymbol n}_{ij})
    (\hat{\boldsymbol k}\cdot \hat{\boldsymbol n}_{\ell m})
    {\cal R}_{ij} {\cal R}_{\ell m}^*
    e^{i \boldsymbol k \cdot \boldsymbol L_{i\ell}}
    = &
    \Bigg[
    \bigg(
    (\hat{\boldsymbol n}_{ij} \cdot \hat{\boldsymbol n}_{\ell m})
    \frac{j_1(\rho_{i\ell})}{\rho_{i\ell}}
    - ( \hat{\boldsymbol n}_{ij} \cdot \hat{\boldsymbol n}_{i \ell} )
    ( \hat{\boldsymbol n}_{\ell m} \cdot \hat{\boldsymbol n}_{i \ell} ) j_2(\rho_{i\ell})
    \bigg)
    \nonumber\\
    & - e^{i \omega L_{ij}} 
    \bigg(
        (\hat{\boldsymbol n}_{ij} \cdot \hat{\boldsymbol n}_{\ell m}) \frac{j_1(\rho_{j\ell})}{\rho_{j\ell}}
        - ( \hat{\boldsymbol n}_{ij} \cdot \hat{\boldsymbol n}_{j \ell} )
    ( \hat{\boldsymbol n}_{\ell m} \cdot \hat{\boldsymbol n}_{j \ell} ) j_2(\rho_{j\ell})
    \bigg)
    \nonumber\\
    & - e^{-i \omega L_{\ell m}} 
    \bigg(
        (\hat{\boldsymbol n}_{ij} \cdot \hat{\boldsymbol n}_{\ell m}) \frac{j_1(\rho_{im})}{\rho_{im}}
        - ( \hat{\boldsymbol n}_{ij} \cdot \hat{\boldsymbol n}_{i m} )
    ( \hat{\boldsymbol n}_{\ell m} \cdot \hat{\boldsymbol n}_{i m} ) j_2(\rho_{im})
    \bigg)
    \nonumber\\
    & + e^{i \omega(L_{ij} - L_{\ell m})} 
    \bigg(
        (\hat{\boldsymbol n}_{ij} \cdot \hat{\boldsymbol n}_{\ell m})  \frac{j_1(\rho_{jm})}{\rho_{jm}}
        - ( \hat{\boldsymbol n}_{ij} \cdot \hat{\boldsymbol n}_{jm} ) ( \hat{\boldsymbol n}_{\ell m} \cdot \hat{\boldsymbol n}_{jm} ) j_2(\rho_{jm})
    \bigg)
    \Bigg]\,,
    \label{angular_integral}
\end{align}
\end{widetext}
where $j_n(x)$ is a spherical Bessel function of the first kind, and $\rho_{ij} = |\boldsymbol k| L_{ij} = x m_\phi \sigma L_{ij}$. 

\subsubsection{Time-Delay Interferometry}
In this subsection, we present the computation of the power spectrum of the TDI variables, Eqs.~\eqref{SXX_ULDM}--\eqref{SXY_ULDM}. Since TDI variables can be written as a linear combination of single-link observables $s_{ij}$, the spectrum and the response integral of any TDI variable can be constructed by combining the cross spectral density with appropriate weights. 

Instead, we start by deriving the Fourier expression of the TDI variables and compute the power spectrum directly from such an expression. In the time-domain, the Michelson TDI variables are defined as~\cite{armstrongTimeDelayInterferometrySpacebased1999, Shaddock:2003dj}
\begin{align}
    {\rm X}
    =&
    \big[ s_{13} + D_{13} s_{31} + D_{13} D_{31} s_{12} + D_{13} D_{31} D_{12} s_{21} \big]
    \nonumber\\
    & - \big[ s_{12} + D_{12} s_{21} + D_{12} D_{21} s_{13} + D_{12} D_{21} D_{13} s_{31} \big] , 
    \nonumber \\
    {\rm Y}
    =&
    \big[ s_{21} + D_{21} s_{12} + D_{21} D_{12} s_{23} + D_{21} D_{12} D_{23} s_{32} \big]
    \nonumber\\
    & - \big[ s_{23} + D_{23} s_{32} + D_{23} D_{32} s_{21} + D_{23} D_{32} D_{21} s_{12} \big] , 
    \nonumber \\
    {\rm Z}
    =&
    \big[ s_{32} + D_{32} s_{23} + D_{32} D_{23} s_{31} + D_{32} D_{23} D_{31} s_{13} \big]
    \nonumber\\
    & - \big[ s_{31} + D_{31} s_{13} + D_{31} D_{13} s_{32} + D_{31} D_{13} D_{32} s_{23} \big] , 
    \nonumber
\end{align}
where $D_{ij} f(t) = f(t - L_{ij})$. As before, we assume $L_{ij}=L$ for simplicity. With this assumption, the TDI X variable in frequency space is
\begin{align}
    \widetilde{\rm X}
    &=
    (1 - \tilde D^2) 
    \Big[ 
        \big( \tilde s_{13} - \tilde s_{12} \big)
        + \tilde D \big( \tilde s_{31} - \tilde s_{21} \big)
    \Big]
    \nonumber \\
    &= \frac{i\omega_L}{\omega^2}
    \int \frac{d^3k}{(2\pi)^3}
    e^{i \boldsymbol k \cdot \boldsymbol x_1}
    (\boldsymbol k \cdot \boldsymbol R_{\rm X}) \tilde{\cal U}(k)\,,
\end{align}
where we introduce a response function of the TDI variable ${\rm X}$ as
\begin{align}
    \boldsymbol R_{\rm X}
    =& 
    2 \tilde D (1  - \tilde D^2) 
    \Big[
    \hat{\boldsymbol n}_{31} \big(e^{i \boldsymbol k \cdot \boldsymbol L_{31}} - \cos (\omega L) \big) 
    \nonumber\\
    & 
    - \hat{\boldsymbol n}_{21} \big(e^{i \boldsymbol k \cdot \boldsymbol L_{21}} - \cos (\omega L) \big) 
    \Big]\,,
\end{align}
Here we use $\tilde D = e^{i\omega L}$ and $\tilde s_{ij}(f)$ given in Eq.~\eqref{sij_freq}. Similar expressions for Y and Z can be obtained by cyclical rotation of the spacecraft indices $1\to 2 \to 3 \to 1$.  

The cross-power spectral density of TDI variables can be then decomposed in the same way as in \eqref{single_link_decomposition}:
\begin{align}
    S_{UV}(f) = 
    S_{\delta}(f)
    {\cal I}_{UV}(f)\,,
\end{align}
where $S_\delta(f)$ is given in Eq.~\eqref{S_DM}, and the response integral of TDI variables is defined similarly as
\begin{align}
    {\cal I}_{UV}
    &\equiv 
    3 \int_0^\infty dx \, p(x) 
    \int \frac{d\Omega}{4\pi}
    (\hat{\boldsymbol k} \cdot \boldsymbol R_U)
    (\hat{\boldsymbol k} \cdot \boldsymbol R_V^*)
    e^{i \boldsymbol k \cdot \boldsymbol L_{UV}} \,. 
\end{align}
Here $\boldsymbol L_{UV} = \boldsymbol x_U - \boldsymbol x_V$ and $\boldsymbol x_{U,V}$ is an anchor position of Michelson TDI variables; for the XYZ variables, we choose $\boldsymbol x_{1,2,3}$ as their anchor position. Here $\boldsymbol x_{1,2,3}$ is the position of the test mass loaded in spacecraft $1,2,3$. Under the assumption $L_{ij} = L$, the correlator of XYZ variables are completely characterized by the common diagonal and off-diagonal parts since $S_{\rm{XX}} = S_{\rm{YY}} = S_{\rm{ZZ}}$ and $S_{\rm{XY}} = S_{\rm{YZ}} = S_{\rm{ZX}}$. Therefore, below, we only compute the angular integral for ${\cal I}_{\rm{XX}}$ and ${\cal I}_{\rm{XY}}$.

Using the identity \eqref{identity}, we find the angular part of the integral as
\begin{widetext}
\begin{align}
    &
    \int \frac{d\Omega}{4\pi}
    (\hat{\boldsymbol k} \cdot \boldsymbol R_{\rm X})
    (\hat{\boldsymbol k} \cdot \boldsymbol R_{\rm X}^*)
    \nonumber\\
    &= \frac{16\sin^2\omega L(2+\cos^2(\omega L))}{3}
    \bigg[
    1 - \frac{3}{2 + \cos^2(\omega L)}
    \bigg(
    \bigg(
        \frac{j_1(\rho)}{\rho}
        + \frac{j_2(\rho)}{2}
    \bigg)
    + 2 \cos(\omega L)
    \bigg(
        \frac{j_1(\rho)}{\rho}
        - j_2(\rho)
    \bigg)
    \bigg)
    \bigg] , 
    \\
    & \int \frac{d\Omega}{4\pi}
    (\hat{\boldsymbol k} \cdot \boldsymbol R_{\rm X})
    (\hat{\boldsymbol k} \cdot \boldsymbol R_{\rm Y}^*) e^{i \boldsymbol k \cdot \boldsymbol L_{12}}
    \nonumber\\
    &= 
    - \frac{16 \sin^2(\omega L) (1 + 2 \cos \omega L)}{6}
    \bigg[
    1 - \frac{3\cos\omega L}{1 + 2 \cos \omega L}
    \bigg(
        \bigg( \frac{4j_1(\rho)}{\rho} - j_2(\rho) \bigg)
        - \cos \omega L \bigg( \frac{j_1(\rho)}{\rho} + \frac{j_2(\rho)}{2} \bigg)
    \bigg)
    \bigg]. 
\end{align}
\end{widetext}
For these expressions, we assumed that the angle between each arm is $60^\circ$. By factorizing sinusoidal factors, we reproduce the power spectrum and the response integral of TDI variables given in Eqs.~\eqref{SXX_ULDM}--\eqref{IXY}. 

\bibliographystyle{utphys}
\bibliography{ref}
\end{document}